\documentstyle[12pt,graphicx,color,subfigure,cite]{article}
\newcommand{\comment}[1]{}
\newcommand{\newc}{\newcommand}

\def\issue(#1,#2,#3){{\bf #1}, #2 (#3)}

\def\PREP(#1,#2,#3){Phys.\ Rep. \issue(#1,#2,#3)}
\def\EPJC(#1,#2,#3){Eur.\ Phys.\ J.\ C \issue(#1,#2,#3)}

\def\tanb{\tan\beta}

\def\thetab0{\theta_{B_0}}

\def\r2{\sqrt 2}
\def\beq{\begin{equation}}
\def\eeq{\end{equation}}
\def\beqn{\begin{eqnarray}}
\def\eeqn{\end{eqnarray}}
\def\sinW2{\sin^2\theta_W}

\def\mz2{M_{z}^2}
\def\c2b{\cos 2\beta}

\def\mx#1{m_{{\tilde \chi}^{0}_#1}}

\def\mz{M_Z}
\def\mhalf{m_{\frac{1}{2}}}

\def\sec2w{sec^2\theta_W}

\def\gmin2{(g-2)_\mu}

\catcode`\@=11 

\def\lsim{\mathrel{\mathpalette\@versim<}}
\def\gsim{\mathrel{\mathpalette\@versim>}}
\def\@versim#1#2{\vcenter{\offinterlineskip
    \ialign{$\m@th#1\hfil##\hfil$\crcr#2\crcr\sim\crcr } }}

\def\te{\tilde e}

\def\tu{\tilde u}

\def\tb{\tilde b}

\def\tst{\tilde t}

\def\tg{\tilde g}

\def\tz{\widetilde Z}

\newc{\wt}{\widetilde}
\newc{\ra}{\rightarrow}
\newc{\s}{\smallskip}
\newc{\nn}{\noindent}
\newc{\non}{\nonumber}

\def \chonep{{\wt\chi_1}^{+}}
\def \chonem{{\wt\chi_1^-}}
\def \chonep2{{\wt\chi_2^+}}
\def \chonem2{{\wt\chi_2^-}}

\def \lspone{\wt\chi_1^0}
\def \mlspone{m_{\lspone}}
\def \lsptwo{\wt\chi_2^0}
\def \mlsptwo{m_{\lsptwo}}
\def \lspthree{\wt\chi_3^0}
\def \mlspthree{m_{\lspthree}}
\def \lspfour{\wt\chi_4^0}
\def \mlspfour{m_{\lspfour}}

\def\NP{Nucl. Phys.}

\def\mygraph#1#2{ \subfigure[]{
   \label{#1}
   \hspace*{-0.6in}
   \begin{minipage}[b]{0.5\textwidth}
   \centering
   \hspace*{4ex}
   \includegraphics[width=0.9\textwidth,height=0.7\textwidth]{#2}
   \vspace*{-4ex}
   \end{minipage}}
   \vspace*{-1ex}
}

\begin{document}
\begin{center}
  { \Large\bf 
 Higgs funnel region of SUSY 
dark matter for small $\tan\beta$ and renormalization group 
effects on pseudoscalar Higgs boson 
with scalar mass non-universality 
\\}
  \vglue 0.5cm
  Utpal Chattopadhyay and Debottam Das\footnote{Emails: tpuc@iacs.res.in, 
tpdd@iacs.res.in}
    \vglue 0.2cm
    {\em Department of Theoretical Physics\\ Indian Association 
for the Cultivation of Science\\ 
2A and 2B Raja S.C. Mullick Road\\
Jadavpur, Kolkata 700 032, India}
  \end{center}        
\begin{abstract}
A non-universal scalar mass supergravity type of model is 
explored where the first two generation of scalars and the 
third generation of sleptons may be very massive.  The lighter 
or vanishing third generation  of  squarks as well as Higgs  
scalars at the unification scale cause  the radiative 
electroweak  symmetry breaking constraint to be less  prohibitive.    
Thus, both FCNC/CP-violation problems as well as the naturalness 
problem are within control.  We identify a large slepton 
mass effect in the RGE  of $m_{H_D}^2$ (for the down type of 
Higgs) that may turn the later negative at the electroweak 
scale even for a small $\tan\beta$.  A hyperbolic branch/focus 
point like effect is found for $m_A^2$ that may result in very light 
Higgs spectra.  The  lightest stable  particle is dominantly a 
bino that pair annihilates via Higgs exchange, giving rise
to  a  WMAP  satisfied  relic  density region for all  
$\tan\beta$. Detection  prospects  of  such  LSPs in  the
upcoming  dark matter experiments  both of  direct and  indirect types
(photon flux) are interesting.  The  Higgs bosons and  the third 
generation of  squarks are light in  this scenario and these 
may be  easily probed besides charginos and neutralinos in 
the early runs of LHC.
\end{abstract}
\section{Introduction}
\label{introduction}
 Low energy supersymmetry (SUSY)\cite{SUSY}
 is one of the most active fields of research 
for physics beyond the standard model (SM)\cite{standardmodel}. 
A minimal extension of the 
Standard Model when supersymmetry is incorporated is the Minimal 
Supersymmetric Standard Model (MSSM)\cite{KaneKingRev,SUSY}
 that includes two Higgs doublets. 
The model however has a large number of SUSY breaking parameters and this 
motivates one into studying models with specific mechanisms for breaking 
SUSY. The later involves high scale physics input and renormalization group 
analyses. This in general leads to a large reduction of the number of 
unknown parameters.  
The minimal supergravity (mSUGRA)\cite{msugra} 
model is one of the well  
studied SUSY models. It requires a very few input parameters at the 
gauge coupling unification scale or grand unified theory (GUT) scale, 
$M_G \simeq 2\times 10^{16}$~GeV. The parameters indeed quantify 
our ignorance of the exact 
nature of SUSY breaking. 
The model incorporates radiative breaking of 
electroweak symmetry (REWSB). The unification scale universal 
input parameters are:  
i) the gaugino mass parameter $\mhalf$, ii) the 
scalar mass parameter $m_0$ and iii) the tri-linear SUSY breaking parameter 
$A_0$. Additionally, one has to provide with $\tanb$, 
the ratio of Higgs vacuum expectation values 
and the sign of the Higgsino mixing parameter $\mu$.  
Renormalization group evolutions are used to obtain the electroweak scale 
parameters of MSSM.  

There is however 
no {\em a priori} necessity of considering   
such universalities of parameters. Indeed, it is worthwhile 
to explore scenarios with non-universalities in the scalar
or in the gaugino masses 
at the unification scale\cite{berez,Nath:1997qm,Cerdeno:2004zj,ellis-all,baer-all-non,baer-higgs,biswarup-sourov-my-uc} . 
Non-universal scalar masses may appear because of non-flat K\"ahler 
potential\cite{soni}. 
However, one must be careful to accommodate the stringent  
constraints from phenomena involving flavor changing 
neutral currents (FCNC). Satisfying FCNC constraints demands 
near-degeneracy of the first two generations of scalar masses but the 
requirements on the third generation of scalars as well as the Higgs scalars 
are not so stringent
\cite{FCNC,ArkaniHamed:flavor}. 
Thus it is seen that FCNC constraints as 
well as constraints from CP-violating phases (for example, those arising 
from the electric dipole moments of the electron and neutron) 
may be managed by introducing 
multi-TeV scalar masses for the first two 
generations of scalars\cite{Gabbiani:1996hi}. 
The third generations of scalar masses and the Higgs scalar masses however 
should be adequately light 
in order to satisfy naturalness\cite{BG}. 
There have been several efforts for obtaining the desired features as outlined 
above.
Analyses with Radiatively Generated Inverted Mass Hierarchy Models (RIMH)
were made in Ref.\cite{RIMH,ArkaniHamed:flavor,Baer:2000xa}.  
In Ref.\cite{Baer:2000xa} the authors achieved the above-mentioned 
requirements at the 
electroweak scale 
by using $t-b-\tau$ Yukawa 
unification with special non-universal relationship among 
the scalar masses at $M_G$. 
Here, the Higgs and the third generation 
of scalar masses are rapidly diminished at the 
electroweak scale via RG evolutions. However, the Yukawa unification and the consideration  
of REWSB constrain such models heavily.    
A second realization of the above idea is partially possible 
via the hyperbolic branch (HB)/ focus point (FP)\cite{focus,hyper} 
scenarios where the scalars may become considerably massive (multi-TeV) in a
subset of the typical mSUGRA parameter space satisfying universal 
boundary conditions while fine-tuning\cite{BG} still remains small. 
A third possibility 
was considered in Ref.\cite{Barger:1999iv}, where plain decoupling 
arguments motivated the authors in using explicit splitting at $M_G$ 
between the scalars belonging to the first two generations and the same of the 
third generation (along with the Higgs scalars).  
Here the first two generations of scalars were chosen 
in the multi-TeV domain whereas
the third generation of scalars as well as the Higgs scalars 
were considered to be in the sub-TeV zone. Additionally, a large value of the
tri-linear coupling parameter was chosen in Ref.\cite{Barger:1999iv} 
for the first two-generations.
Universality of scalar masses 
in the first two-generations along with a choice of a different 
scalar mass parameter for the third generation as well 
as the Higgs scalars, or even splitting of squarks and sleptons within the 
third generation itself have also been 
considered in Refs.\cite{Baer:2004xx,King:2006tf,Cerdeno:2001up}. 
In Ref.\cite{King:2006tf}
the authors additionally considered non-universality in 
the gaugino masses and  
analyzed the fine-tuning aspect of computing the relic density of dark 
matter in addition to obtaining 
a parameter zone of the MSSM corresponding to a well tempered 
neutralino\cite{{Arkani-Hamed:2006mb}}. Similarly, analyses 
with only non-universalities in the  Higgs 
scalar may be seen in 
Refs.\cite{ellis-all,baer-higgs,Nath:1997qm,McMullen:2006mj}. A comprehensive 
set of characteristic possibilities for varieties of 
non-universal SUGRA scenarios may be seen in Ref.\cite{pncompLHC}. 

In this analysis we explore a SUGRA scenario with 
non-universal scalar masses that 
i) allows to have very large first two-generation of scalar masses so as to 
obey the FCNC and the CP-violation limits easily ({\em ie.} without 
requiring any ultra-small phases) and that would 
not impose any additional price on fine-tuning, ii)  
spans a large amount of MSSM parameter space satisfying the neutralino 
relic density constraint from WMAP data by not requiring any delicate 
mixing of bino and Higgsinos, so that we would find 
a bino-dominated lightest neutralino for most of the parameter space and,  
iii) satisfies the Higgs mass lower bound from LEP2 data as well as other 
low energy constraints.  
Certainly, a model with 
REWSB and universal scalar mass like mSUGRA 
is not friendly to achieve these  
objectives if we consider a common gaugino mass parameter below a TeV 
or so.  The above phenomenologically 
inspired requirements in combination motivate us    
to introduce non-universality between the third and the 
first two-generation 
of scalars as well as the Higgs scalars. This has to be such 
that the REWSB conditions would not become prohibitive to have a 
multi-TeV first two-generation of scalars. 
Here we would like to point out that we would not be able to satisfy 
the mentioned objectives by considering non-universalities
only in the Higgs scalar masses. 

A simpler and purely phenomenological 
attempt in this direction in a universal gaugino mass framework 
could be to 
generate all the third generation of scalar masses and the Higgs 
scalar masses radiatively, starting from zero values 
at the unification scale, while keeping the first two-generations  
of scalar masses universal (and this may be in the multi-TeV zone).
We point out that starting with a similar range of values for the 
Higgs scalars as with the third generation of scalars at $M_G$ would 
be desirable since this would lead to similar range of mass values for 
all the soft-SUSY breaking terms contributing to REWSB after RG evolutions.
However, if we consider all the 
third generation of scalars at $M_G$ to be light, we would 
find stau ($\tilde \tau$) becoming the lightest stable particle (LSP) 
or even tachyonic at the electroweak scale. On a similar note, we 
remind that such 
an appearance of tachyonic sleptons also arise in the Anomaly 
Mediated Supersymmetry Breaking (AMSB) analyses, 
and it is avoided purely by a phenomenologically inspired way 
of adding an appropriate non-zero mass value to all the scalars 
at the unification scale in the minimal AMSB scenario\cite{mAMSB}. 
Thus with a simple motivation  
of managing with FCNC and CP-violation, 
naturalness as well as the dark matter constraints 
concurrently we consider a non-zero mass value 
for the third generation of sleptons, while having the 
masses of squarks of the third generation and 
the Higgs scalars vanishing at 
$M_G$. Additionally, for convenience 
we set the third generation of 
slepton mass parameters the same as that of 
the first two-generation of scalars at $M_G$.   

Thus, the parameters of our Non-universal Scalar Mass model which will henceforth be called the NUSM 
model is given by 
\beq
\tan\beta, \mhalf, A_0, sign(\mu),~{\rm and}~m_0 
\eeq
where the scalar mass input assignments at $M_G$ are as follows.  \\
i) The unification scale mass parameter 
for the first two-generations of squarks,sleptons and the
third generation of sleptons is $m_0$, where $m_0$ is allowed to span up to 
a very large value.\\
ii) The mass parameters 
for the third generation of squarks and Higgs scalars are set 
to zero. \\
We could have also chosen a non-vanishing value for ii) 
as long as it is sufficiently small.   
We may note here that different values of mass parameters for squarks  
and sleptons at $M_G$
may appear in orbifold models with large threshold 
corrections. This is related to having different modular 
weights associated with the  
squarks and the sleptons in a given 
generation\cite{Brignole:1993dj,Cerdeno:2007hn}. 

As we will see below, the NUSM model provides 
with a highly bino-dominated neutralino dark matter 
over almost its full range of 
parameter space. It has an interesting feature of 
having a large funnel region, 
a region of parameter space where the associated 
annihilation channels are characterized by the direct channel
pole $2m_{\tilde \chi_1^0} \simeq m_A, m_H$. 
We note that unlike mSUGRA where 
one finds the funnel region only for a large value of $\tan\beta$, here 
in the NUSM one finds it 
for almost all possible values of $\tan\beta (\gsim 5)$. 
We will also see that the NUSM  is further 
characterized by a lighter $m_A$ or $m_H$ particularly when  
$m_0$ is large 
and this is found even for small values of $\tan\beta (\sim 10)$. 
A small part of the parameter space may be far 
from the decoupling\cite{decouplingHiggs} 
region of Higgs boson mixing that causes  
the lighter CP-even Higgs boson ($h$-boson) to be  
non-Standard Model like\cite{nondecouplingHiggs}. 
This in turn reduces the lower bound of $m_h$ much below  
the LEP2 Higgs boson mass limit.

In this work we include 
a semi-analytic calculation that first points out a large mass effect 
in the solution of the renormalization group equations (RGE) 
of $m_{H_D}^2$.  The effect causes a 
hyperbolic branch/focus point like 
behavior\footnote{We remind that 
the well known HB/FP effect that occurs in $m_{H_U}^2$ is associated 
with the first minimization condition of Radiative electroweak symmetry 
breaking. In this analysis we do not have such an effect. On the contrary 
we have a similar HB/FP effect for small values of $\tan\beta$ 
in connection with the 
second minimization condition 
that is associated with $m_A^2$.} 
in $m_A^2 (\simeq m_{H_D}^2 - m_{H_U}^2)$ that may cause  
$m_A$ to become smaller even for a small $\tan\beta$ as mentioned above. 
Here we remind that $m_A$ 
typically becomes smaller in mSUGRA only for large $\tan\beta$.   

  The paper is organized as follows.  In Sec.\ref{nusmhbfp} we will primarily 
discuss the large mass RGE effects in the NUSM  on  
$m_A^2$ and its consequent reduction for large $m_0$ 
via an HB/FP-like effect. We will obtain the   
semi-analytic results which will also be verified by numerical computation. 
In Sec.\ref{cdmandfunnel} we will 
study the cold dark matter constraint from WMAP data including 
also the constraints from the LEP2 Higgs bound, $b \rightarrow s+\gamma$, 
and $ B_s \to \mu^+ \mu^- $. We will also discuss 
a few sample points in the context of LHC reach. In 
Sec.\ref{detectionrates} we will discuss the direct and indirect 
detection (via continuous gamma ray) rates of the LSP. 
Finally we will conclude in Sec.\ref{concl}. 
   
\section{The non-universal scalar scenario:NUSM}
\label{nusmhbfp}
In this section we focus on the key elements that are 
important for the NUSM. 
The primary quantities of phenomenological 
interest are $\mu$, 
and $m_A$.  These in turn have important significance on the 
following:  i) the issue of fine-tuning, 
ii) the presence of non-Standard Model like 
lighter Higgs boson mass bound for a limited region of 
parameter space, iii) dark matter and  
iv) collider discovery possibilities. Additionally, we remind ourselves about 
obtaining a lighter third generation of squarks at the electroweak 
scale and this is of course related to their 
vanishing values at the unification scale.

To start with we write down the REWSB results,  
\beq
\mu^2 = 
-\frac{1}{2} M^2_Z +\frac {m_{H_D}^2-m_{H_U}^2 \tan^2\beta} {\tan^2\beta -1}
+ \frac {\Sigma_1 -\Sigma_2 \tan^2\beta} {\tan^2\beta -1},  
\label{rewsbmusqeqn}
\eeq
and, 
\beq
\sin2\beta =  2B\mu/(m_{H_D}^2+m_{H_U}^2+2\mu^2+\Sigma_1+\Sigma_2)
\label{rewsbBeqn}
\eeq
where $\Sigma_{1,2}$ represent the one-loop 
corrections~\cite{effpot1,effpot2}, that become small in the 
scale where the Higgs potential $V_{Higgs}$ is minimized.
We may approximately consider 
$\mu^2 \simeq -m_{H_U}^2$ (for $\tanb\gsim 5$) and 
$m_A^2=m_{H_D}^2+m_{H_U}^2 +2\mu^2 \simeq m_{H_D}^2-m_{H_U}^2$ at tree level. 

 The associated one-loop RGEs are given by (neglecting the very small 
first two-generations of Yukawa contributions),  
\begin{eqnarray}
{\frac{{dm_{H_D}^2}}{{dt}}} &=&(3\tilde{\alpha}_2{\tilde{m}_2}^2+\frac 35
\tilde{\alpha}_1{\tilde{m}_1}^2)-3Y_b(m_{H_D}^2+
m_Q^2+m_D^2+{A_b}^2) \nonumber \\
&&-Y_\tau (m_{H_D}^2+m_L^2+m_E^2+{A_\tau }^2)+\frac 3{10}\tilde
{\alpha}_1S_0  
\label{Hdeqn}
\end{eqnarray}

\begin{equation}
{\frac{{dm_{H_U}^2}}{{dt}}}=(3\tilde \alpha_2 {\tilde m_2}^2 + \frac{3}{5}
\tilde \alpha_1 {\tilde m_1}^2) -3Y_t(m_{H_U}^2+m_Q^2+m_U^2+{A_t}^2) 
-
\frac{3}{10}\tilde \alpha_1 S_0
\label{Hueqn}
\end{equation}
Here we have $t=ln(M_G^2/Q^2)$ with $Q$ being the renormalization scale.  $\tilde
\alpha_i=\alpha_i/(4\pi)$ for $i=1,2,3$ are the scaled gauge coupling
constants (with $\alpha_1=\frac53 \alpha_Y$) and $\tilde m_i$ 
are the running gaugino masses, where $i=1,2,3$ refers to U(1), SU(2) and 
SU(3) gauge groups respectively. $Y_j$ represent
the scaled and squared Yukawa couplings, 
e.g, $Y_j \equiv h_j^2/(4\pi)^2$ where
$h_j$ is a Yukawa coupling ($j=1,2,3$ stands for $t,b,\tau$ respectively). 
The quantity $S_0$ which is shown below may become relevant 
for a general set of non-universal boundary condition for 
scalars. However in this analysis 
it is zero at $M_G$ and it causes only a negligible effect. 
\beq
S_0=m_{H_U}^2-m_{H_D}^2
+\Sigma_k (m_{q_k}^2 +
m_{d_k}^2 +
m_{e_k}^2 -
m_{l_k}^2-2 m_{\tilde u_k}^2)+(m_Q^2+m_D^2+m_E^2-m_L^2-
     2 m_U^2)
\eeq
where the subscript $k$ for $k=1$ or $2$ indicates the first two generations.  

\noindent
The solutions for $\mu^2$ and $m_A^2$ valid up to a moderately large  
value of $\tanb$ (10 or so) follow from Ref.\cite{Nath:1997qm} 
after appropriately considering the NUSM parameters.  
This involves ignoring $h_b$ and $h_\tau$ 
in the limit of a small $\tan\beta$. The solutions read
\beq
\mu^2=m_0^2C_1 + A_0^2 C_2 +\mhalf^2C_3+\mhalf A_0 C_4-\frac{1}{2}M_Z^2+
\frac{3}{5} \frac{\tan^2\beta+1}{\tan^2\beta-1} S_0 p, \quad {\rm and}
\label{mueqn}
\eeq
\beq
m_A^2=m_0^2D_1 + A_0^2 D_2 +\mhalf^2D_3+\mhalf A_0 D_4-\frac{1}{2}M_Z^2+
\frac{6}{5} \frac{\tan^2\beta+1}{\tan^2\beta-1} S_0 p.
\label{mAeqn}
\eeq
Here $p$ is given by: 
$p=\frac{5}{66}[1-(\frac{\tilde\alpha_1(t)} {\tilde\alpha_1(0)})]$. 
The unification scale conditions for scalar mass squares 
for the NUSM may be given as $m_i^2=(1+\delta_i)m_0^2$. 
Here, the subscript $i$ for $\delta_i$ stands for: 
$i\equiv H_D,H_U,Q_L,u_R,d_R,l_L,e_R$.  For NUSM one has 
$\delta_{H_u}=\delta_{H_D}=\delta_{Q_L}=\delta_{u_R}=\delta_{d_R}=-1$
and $\delta_{l_L}=\delta_{e_R}=0$.  These unification scale 
conditions allow us to define a quantity $\delta$ that appears in the 
expressions of the coefficients $C_1$ and $D_1$ as shown below. It follows  
that $\delta=-1$ for the NUSM, and $\delta=0$ for mSUGRA. We note that 
only $C_1$ among $C_i$'s and $D_1$ among $D_i$'s depend on $\delta$ and 
one obtains\footnote{In 
Ref.\cite{Nath:1997qm} the authors considered non-universalities 
in the masses of Higgs scalars and the third generation scalars 
$m_{Q_L}$ and $m_{u_R}$. However, NUSM additionally 
requires non-universality in $m_{d_R}$. }, \\
\beqn
C_1&=&(1+\delta) \frac{1}{\tan^2\beta-1}
(1-\frac{3D_0-1}{2}\tan^2\beta), \quad {\rm and} \nonumber \\ 
D_1&=&\frac{3}{2}(1+\delta)\frac{\tan^2\beta+1}{\tan^2\beta-1}(1-D_0).
\label{CDeqn}
\eeqn
Here, $D_0\simeq 1-{(m_t/200 \sin\beta)}^2 \lsim 0.2$. 
The definitions of the quantities $C_i$ and $D_i$ for $i\neq 1$ 
may be seen in the appendix.
As mentioned above, for the NUSM one has 
$\delta=-1$. Hence, Eqs.~\ref{mueqn}-\ref{CDeqn} 
show that at the level of 
approximation where $h_b$ and $h_\tau$ are ignored, $\mu^2$ and $m_A^2$ are 
independent of $m_0$.
 We will however see that   
both $\mu^2$ and $m_A^2$ would actually depend on $m_0$. While the 
dependence of $\mu^2$ on $m_0$ would be caused by two-loop RGE 
effects in $m_{H_U}^2$, the same for $m_A^2$ is very prominent 
even at the one-loop level of RGE and this is found when we include 
the effects due to $h_b$ and $h_\tau$ which are specially important 
for the NUSM. The following subsection 
describes our improved results. 
\subsection{Large slepton mass RGE effect on pseudoscalar Higgs 
boson-- A hyperbolic branch/focus point like effect in $m_A^2$ for small 
$\tan\beta$}
Numerical computation shows that unlike what is seen from Eqs.~\ref{mAeqn} 
and~\ref{CDeqn}, $m_A^2$ 
may indeed decrease very rapidly with an increase of 
$m_0$ for large values of the later. 
There are two reasons behind the above behavior: 
i) the choice of vanishing Higgs scalars in the NUSM at the unification scale 
$M_G$ and ii) a large slepton mass (LSM) effect in the RGE of 
Eq.\ref{Hdeqn} via the tau-Yukawa term. Henceforth, we will refer the combined 
effect of i) and ii) as the LSM effect.  In this analysis we first 
compute $m_{H_D}^2$ analytically without ignoring the terms involving 
$h_b$ and $h_\tau$. The calculation essentially keeps 
the terms involving the bottom and the 
tau-Yukawa couplings similar to 
what was used for the top-Yukawa term of Eq.\ref{Hdeqn} 
in Refs.\cite{Ibanez:1984-85,Nath:1997qm}. This results into, 
\beq
m_{H_{D}}^{2}=C_{H_D}m_{0}^{2}+m_{1/2}^{2} g(t)+\frac{3}{5}S_0p.
\label{mhdsoln}
\eeq
Here $g(t)$ is a function of $\tan\beta$~\cite{Ibanez:1984-85}.  
The term involving $S_0$ vanishes in the NUSM. 
At the level of approximation where the bottom and tau Yukawa couplings are 
straightway ignored in Eq.\ref{Hdeqn}, $C_{H_D}$ would become zero in the 
NUSM. 
The result of our computation of $C_{H_D}$ that leads to a non-vanishing 
value is given below. 
\beq
C_{H_D}=C_{H_D}(U)+C_{H_D}(NU)
\eeq
where,
\begin{eqnarray}
C_{H_D}(U)&=& 1-3(3I_2 +I_3)\nonumber \\
C_{H_D}(NU)&=&\delta_{H_D}-3I_2(\delta_{Q_L}+\delta_{d_R}+\delta_{H_D})
\nonumber \\
&-& I_3(\delta_{l_L}+\delta_{e_R}+\delta_{H_D}).
\label{chdeqn}
\end{eqnarray}
The quantities 
$I_2$ and $I_3$ are functions of 
$Y_i={{h_i^2} \over {{(4\pi)}^2}}$ for $i=2,3~(b,\tau)$. Small $\tan\beta$ 
solutions of $Y_i$s, computed at the electroweak scale\cite{Ibanez:1984-85} 
are shown in the appendix.  
$I_2$ and $I_3$ defined below are computed numerically. 
\beq
I_2=\int_0^t Y_2(t^\prime)dt^\prime,~~I_3=\int_0^t Y_3
(t^\prime)dt^\prime  
\eeq
Here $t=ln(M_G^2/M_Z^2)$. One finds $I_2, I_3 <<1$. 
For NUSM, the above reduces to
\beq
C_{H_D}=-2I_3.
\label{chdneweqn}
\eeq
Since $I_3$ is proportional to the square of $\tau$-Yukawa coupling 
$h_\tau$ where $h_\tau \propto \frac{1}{\cos\beta}$, we find  
$|C_{H_D}|$ to be an increasing function of $\tan\beta$. 
For $\tan\beta=10$, we find 
$C_{H_D}\simeq-0.007$. The above clearly shows the large slepton mass 
effect because the dependence of $C_{H_D}$ or $m_{H_D}^2$ on $I_3$  
arises from the term of Eq.\ref{Hdeqn} that is associated with 
$h_\tau^2$.   
Thus with the NUSM parameters, the scalar mass term that appears as the first 
term in Eq.\ref{mhdsoln} is essentially contributed by the third generation 
of slepton masses.   
Eq.\ref{mhdsoln} and Eq.\ref{chdneweqn} clearly show that 
even with a small $\tan\beta$, 
larger values of $m_0$ 
may reduce $m_{H_D}^2$ appreciably and 
may turn the later negative. This demonstrates 
the LSM effect as mentioned 
before\footnote{(i)We must mention that if we had chosen 
the values of Higgs scalar masses to be $m_0$ at the unification scale, the 
resulting coefficient of $m_0^2$ in $m_{H_D}^2$ would be positive and the 
later would no longer be a decreasing function of $m_0$. 
(ii)~We further note that the fact that the third generation of 
squark masses are vanishing in NUSM also makes the LSM effect prominent. 
For example, had we considered non-vanishing values  
for $m_Q$ and $m_D$ ($=m_0$, ie. $\delta_{Q_L}=\delta_{d_R}=0$) 
we would find the terms in Eq.\ref{chdeqn} that are associated with $I_2$ 
to become more prominent. This would have caused 
$m_{H_D}^2$ to be further negative. However, it 
could also cause $m_Q^2$ to turn 
tachyonic at the electroweak scale. We have also checked this fact 
numerically. 
}.  
We are not aware of any past reference that pointed out 
this large mass effect or provided with semi-analytic expressions. 
For larger $\tan\beta$, $C_{H_D}$ may be appreciably large 
and negative that causes the LSM effect to become prominent even for 
a smaller value of $m_0$. 

  We note that there is no LSM effect that may modify 
$\mu^2$ of Eq.\ref{mueqn} at the one-loop level. 
This is simply because $\mu^2 \simeq -m_{H_U}^2$ 
and there is no possibility of having a LSM effect in 
the corresponding RGE of Eq.\ref{Hueqn}. 
On the other hand, an improved $m_{H_D}^2$ as obtained above in 
Eqs.\ref{mhdsoln} and \ref{chdneweqn}
changes the result of $m_A^2$ (Eq.\ref{mAeqn} and \ref{CDeqn}) 
but the later does not receive 
any additional $m_0^2$ dependence other than what is already contributed 
from Eq.\ref{chdneweqn}.  Thus it follows that the coefficient of $m_0^2$ when 
Eq.\ref{mAeqn} is modified would be $C_{H_D}$.  
As a result, in the above analysis that is valid for small 
$\tan\beta$ below 10 or 15, 
we find that the LSM effect causes $m_A^2$ to have   
a HB/FP like behavior for 
its dependence on $m_0$ and $\mhalf$. Hence, $m_A$ may become 
significantly small for a large $m_0$ because of a cancellation between 
the terms. This of course may happen even for a small value of 
$\tan\beta$. A very large 
$m_0$ would cause $m_A^2$ to become tachyonic, or would result 
into an absence of 
REWSB (Eq.\ref{rewsbBeqn}).  For a large $\tan\beta$ on the other hand, 
the LSM effect is drastically enhanced. We comment 
here that in spite of showing the one-loop 
results we performed a complete numerical solution 
of the RGEs up to two loops in this analysis using SUSPECT\cite{suspect}.    

We now point out that a simple non-universal Higgs scalar 
mass scenario as described in Ref.\cite{baer-higgs} may also provide 
a small $m_A$ along with a funnel type of
region that satisfies the WMAP data, for a smaller value of $\tan\beta$.
We emphasize in particular the case 
where non-universality was analyzed with a unified Higgs scalar 
mass in Ref.\cite{baer-higgs}.  
However, unlike the NUSM these scenarios are 
very much constrained via REWSB so that considering 
larger values of masses for the 
first two-generation of scalars, an easier way to control FCNC and 
CP-violation effects is not possible. 
In these non-universal Higgs scalar models 
$m_A$ becomes small for small values of $\tan\beta$,  
only when tachyonic values of 
$m_{H_D}^2$  and $m_{H_U}^2$ (assumed equal) are considered 
at the unification scale and this severely reduces the available 
parameter space. 
Additionally, one may obtain an $A$-pole annihilation region or 
a funnel region for a small value of $\tan\beta$ 
in the so called sub-GUT CMSSM scenario\cite{Ellis:2006vc}.   

\begin{figure}[!htb]
\vspace*{-0.05in}
\mygraph{m0onmuandmaA}{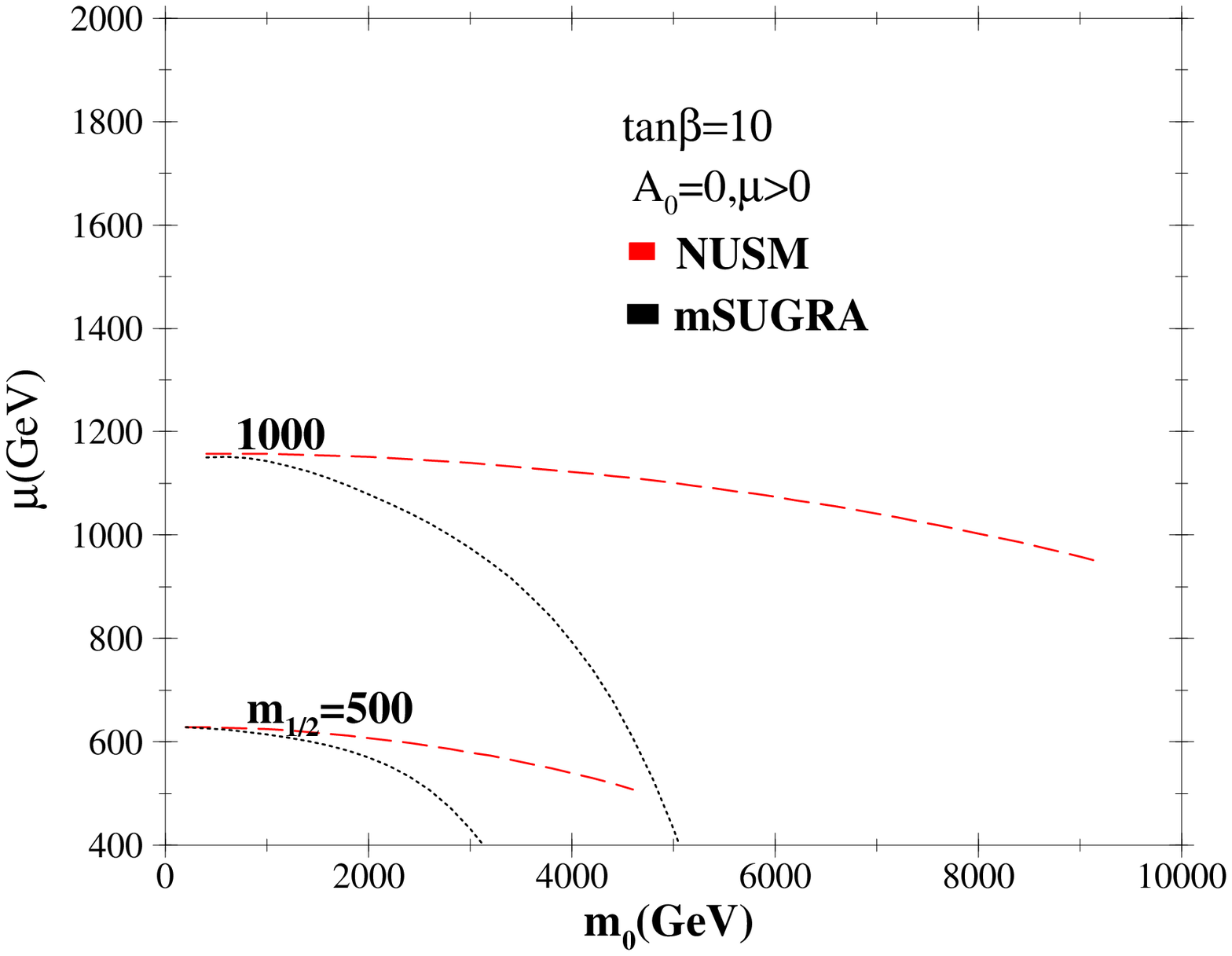}
\hspace*{0.5in}
\mygraph{m0onmuandmaB}{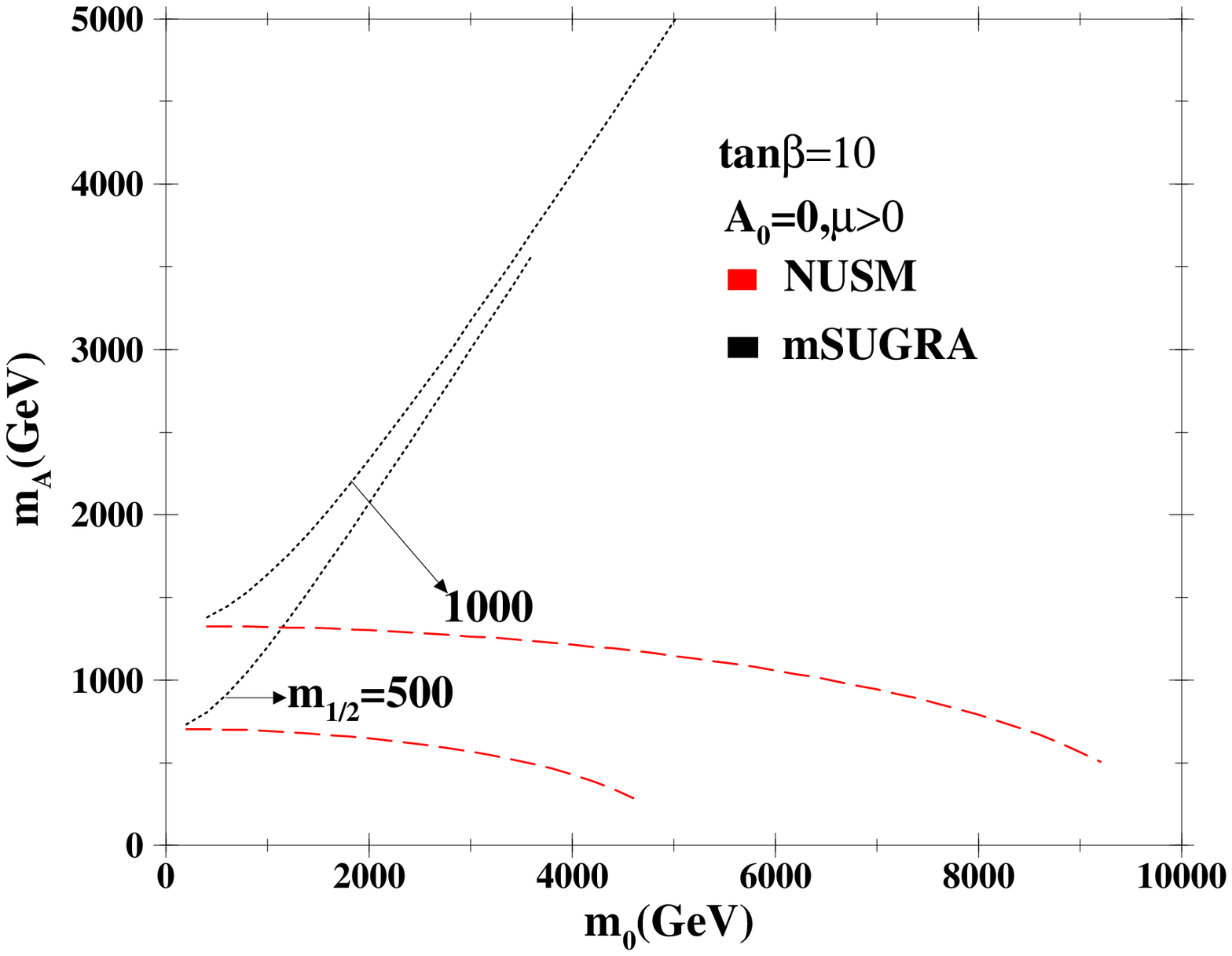}

\vspace*{0.5cm}
\hspace*{1.5in}
\mygraph{m0onmuandmaC}{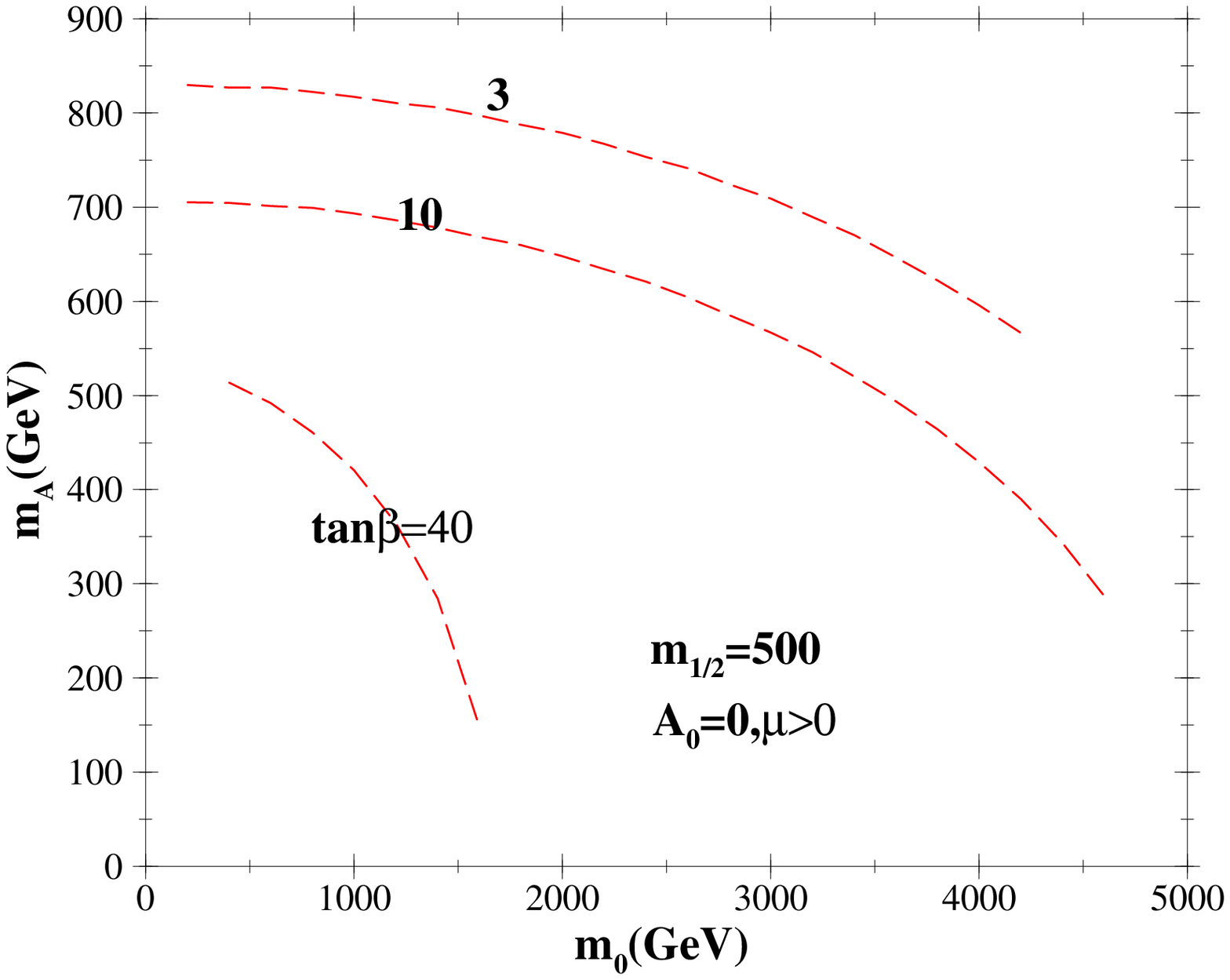}
\caption{
~(a)Variation of $\mu$ with $m_0$ for $\tanb=10$, $A_0=0$ and 
$\mhalf=500,1000$GeV. 
The red dashed lines represent the curves for the NUSM and the black dotted
lines represent the same for mSUGRA.  
~(b) Same as (a), but CP-odd Higgs ($m_A$) is plotted against $m_0$.~(c) 
shows the variation of $m_A$ with $m_0$ in the NUSM  
for different values of $\tanb$ when $\mhalf=500$GeV.  
}
\label{m0onmuandma}
\end{figure}

    We will now describe our results as obtained by using 
SUSPECT\cite{suspect}. 
Fig.\ref{m0onmuandma} shows the 
variation of $\mu$  and $m_A$ when $m_0$ is varied in the NUSM and mSUGRA. 
Fig.\ref{m0onmuandmaA} shows the effect 
of varying $m_0$ in mSUGRA and in the NUSM 
for $m_{1/2}=500$~GeV and 1 TeV. A small 
reduction of $|\mu|$ ($(10 ~{\rm to}~20 \%)$) is seen in the NUSM 
when $m_0$ is increased up to 5 or 10 TeV.  The approximate 
one-loop result of Eq.\ref{mueqn} and Eq.\ref{CDeqn} with $\delta=-1$ as in  
NUSM however would indicate a flat 
$\mu$ over a variation of $m_0$. The figure of course shows a moderately 
varying $\mu$ and we have checked that this variation has its origin  
in two-loop RGE effects. The two different $m_{1/2}$  
contours for mSUGRA however show a decreasing behavior of 
$\mu$ when $m_0$ is enhanced. This happens simply because of the HB/FP  
effect existing in mSUGRA. 
Fig.\ref{m0onmuandmaB} shows the results for $m_A$ which have 
two sets of contours for $m_{1/2}=500$~GeV and 1 TeV corresponding to 
mSUGRA and the NUSM.
In contrast to 
mSUGRA where $m_A$ rapidly increases with $m_0$, the behavior in NUSM is 
opposite so that we may find a very light pseudoscalar
Higgs boson via the LSM effect.  The decrease of $m_A$ 
is even more pronounced for a large 
value of $\tan\beta$ and this may 
be easily seen in Fig.\ref{m0onmuandmaC} where a variation of $m_A$ 
{\it vs} $m_0$ is shown for a given $m_{1/2}(=500~{\rm GeV})$ for three 
different values of $\tan\beta$. The curves ends in the larger $m_0$ sides for
a few different reasons. For $\tan\beta \lsim 10$, the largest $m_0$ limit 
is caused by stop mass becoming very light or unphysical. On the other hand 
for $\tan\beta \gsim 15$ the largest $m_0$ limit is given by absence of REWSB 
(ie $m_A^2$ turning negative).  

\begin{figure}[!htb]
\vspace*{-0.05in}
\mygraph{masssqvsQA}{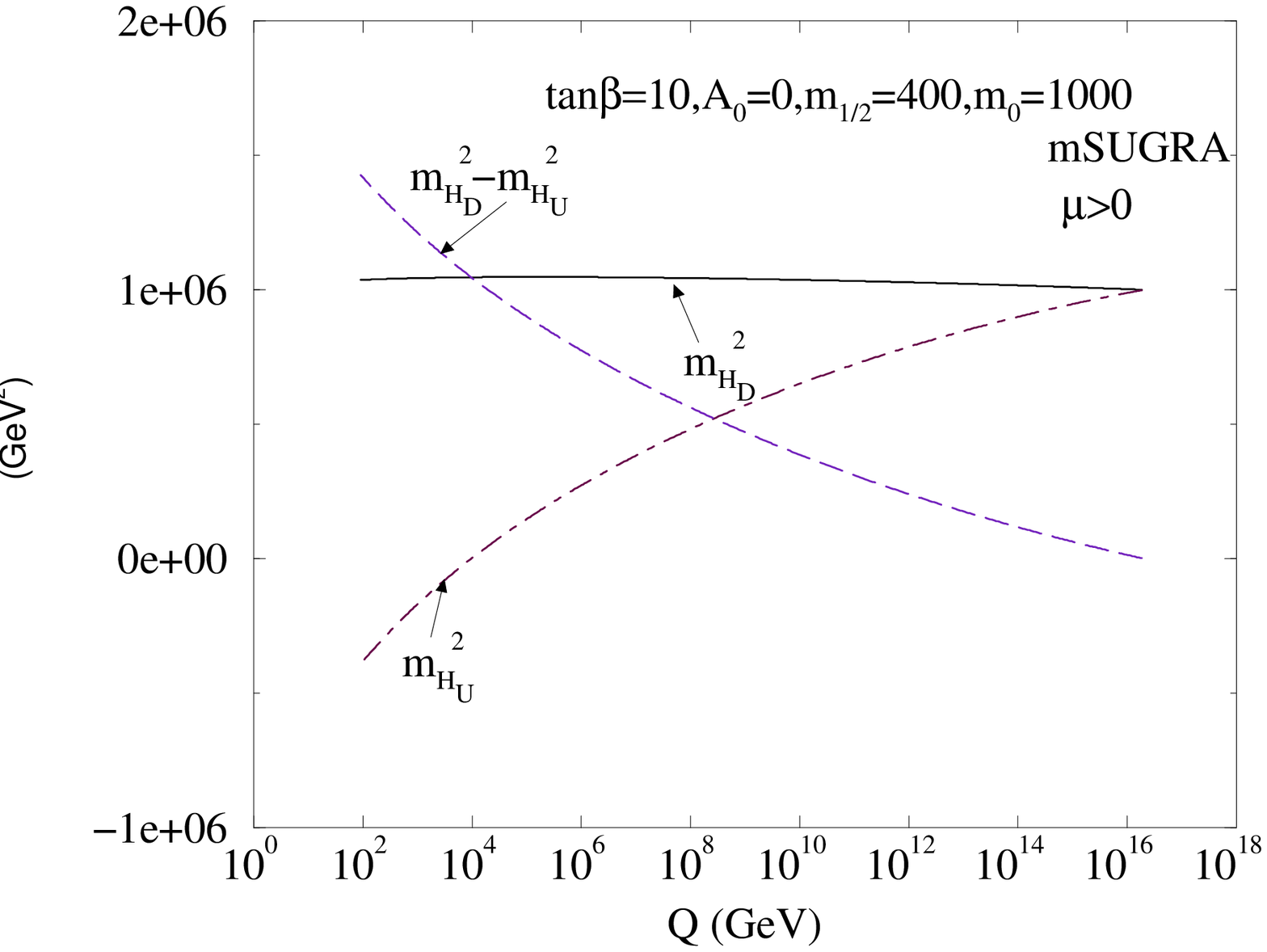}
\hspace*{0.5in}
\mygraph{masssqvsQB}{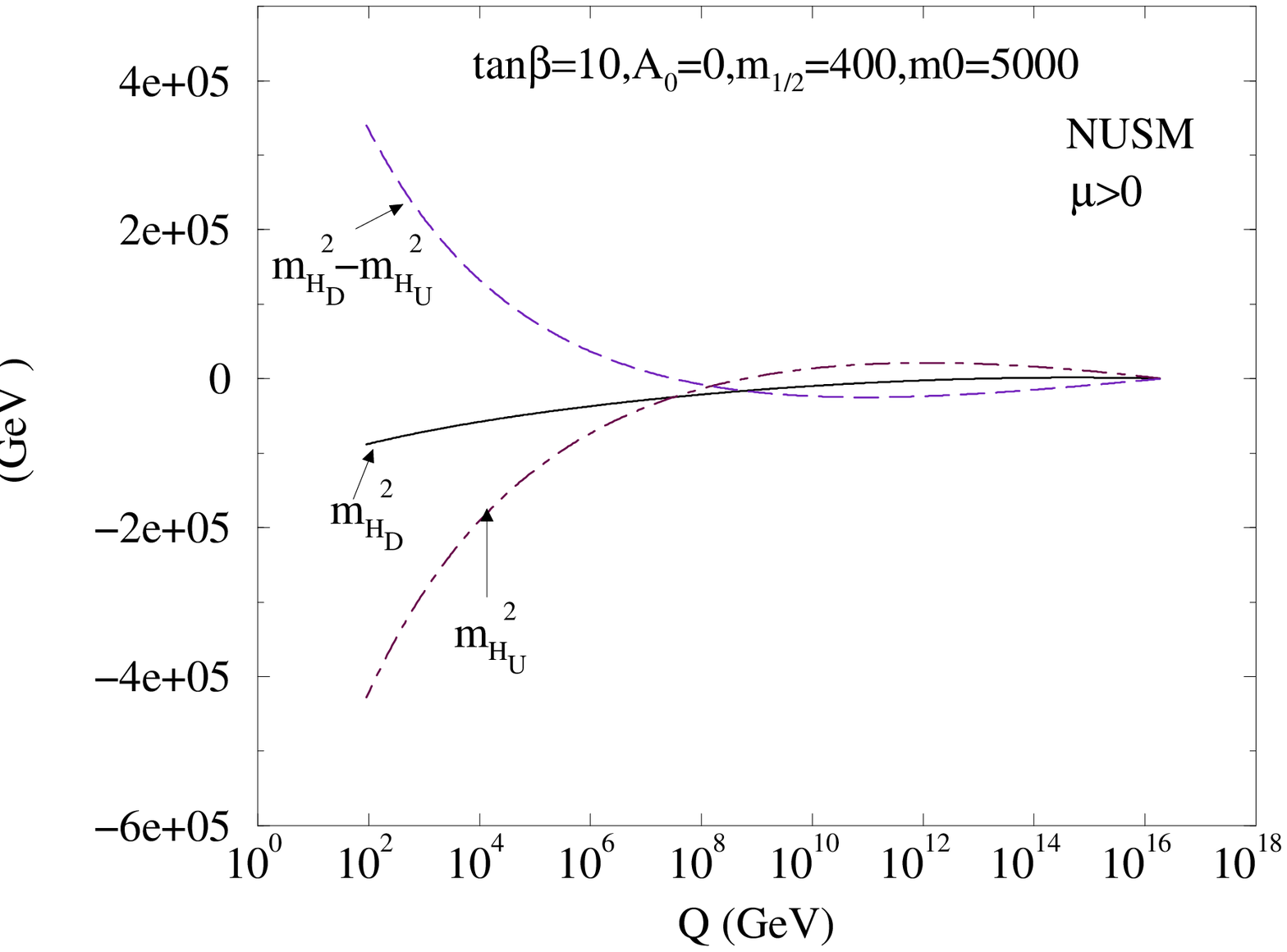}
\caption{
~(a)Variation of squared Higgs scalar masses and their difference 
($\equiv m_A^2 \simeq m_{H_D}^2 - m_{H_U}^2 $)  
with renormalization scale $Q$ 
for $\tanb=10$, $\mhalf=400$~GeV, $m_0=1000$~GeV and $A_0=0$ for 
mSUGRA. 
~(b) Same as (a), except for $m_0=5000$~GeV and for the NUSM . $m_{H_D}^2$ 
turns negative via LSM effect. 
}
\label{masssqvsQ}
\end{figure}

The LSM effect may be explicitly seen in Fig.\ref{masssqvsQ} 
where we plot squared scalar masses of Higgs scalars as well as their 
difference $m_A^2 \simeq m_{H_D}^2 - m_{H_U}^2$ with respect to 
a variation over the renormalization scale $Q$ for mSUGRA and the NUSM  for 
$\tan\beta=10$, $m_{1/2}=400$~GeV and $A_0=0$. 
Fig.\ref{masssqvsQA} corresponds to mSUGRA for $m_0=1$~TeV whereas 
Fig.\ref{masssqvsQB} shows the case of the NUSM  for $m_0=5$~TeV. We have 
chosen different values of $m_0$ in the two figures because of the fact that a 
large $m_0$ is prohibited in mSUGRA via the REWSB constraint whereas 
a choice of a small $m_0$ in the NUSM  would have a negligible 
LSM effect. Clearly, $m_{H_D}^2$ stays positive 
at the electroweak scale in mSUGRA (Fig.\ref{masssqvsQA}).  On the other hand 
the same for the NUSM  turns toward a negative value while running 
from $M_G$ to the electroweak scale (Fig.\ref{masssqvsQB}) and this   
essentially shows the LSM effect in $m_A^2$. Thus a typical 
large $\tan\beta$ phenomenon that occurs in mSUGRA is obtained in the NUSM  
for a small $\tan\beta$.     

\section{Cold dark matter constraint and extended funnel region}
\label{cdmandfunnel}
In supergravity type of models $\lspone$
becomes the LSP for most of the
parameter space\cite{DMreview,recentSUSYDMreview} and we assume that 
the cold dark matter relic density is entirely due to  
$\lspone$. Considering WMAP data\cite{WMAPdata} 
one finds a 3$\sigma$ limit as shown below. 
\begin{equation}
0.091 < \Omega_{CDM}h^2 < 0.128
\label{relicdensity}
\end{equation}
where $\Omega_{CDM}h^2$ is the DM relic density in units of the critical
density. Here, $h=0.71\pm0.026$ is the Hubble constant in units of
$100 \ \rm Km \ \rm s^{-1}\ \rm Mpc^{-1}$.  
In the thermal description, the LSP was in thermal 
equilibrium with the annihilation products at a very high 
temperature of the early universe ($T>>m_{{\tilde \chi}_1^0}$). 
The annihilation products 
include fermion pairs, gauge boson pairs,  
Higgs boson pairs or
gauge boson-Higgs boson combinations and they are produced via 
$s,t$ and $u$-channel processes. As the temperature decreased the annihilation rate
fell below the expansion rate of the universe and the LSP 
went away from the thermal equilibrium and freeze-out occured.  
The current 
value of $\Omega_{{\tilde \chi}_1^0}h^2$ is computed by solving the Boltzmann equation for $n_{{\tilde \chi}_1^0}$, the number
density of the LSP in a Friedmann-Robertson-Walker universe. The above 
computation  
essentially involves finding the thermally averaged quantity 
$<\sigma_{eff} v>$, where
$v$ is the relative velocity between two annihilating neutralinos and
$\sigma_{eff}$ is the neutralino annihilation cross section that includes 
all the final states. In addition to annihilations one considers 
coannihilations\cite{coannistau,coannistop,Edsjo:1997bg,mizuta,
coanniSet2} which are annihilations of the LSP with sparticles 
close in mass values with that of the LSP.  The cross section sensitively 
depends on the nature of the  
composition of the LSP. In the MSSM, the LSP is a mixed state of 
bino ($\tilde B$), wino ($\tilde W$) and 
Higgsinos (${\tilde H}_1^0$ ${\tilde H}_2^0$): 
\begin{equation}
{\tilde \chi}_1^0=N_{11}\tilde B + N_{12}{\tilde W}_3 +
N_{13}{\tilde H}_1^0 +N_{14}{\tilde H}_2^0.
\end{equation}
Here 
$N_{ij}$  are the elements of
the matrix that diagonalizes the neutralino mass matrix. The 
gaugino fraction $F_G$ of the lightest neutralino 
is defined by $F_g=|N_{11}|^2 + |N_{12}|^2$. A gaugino-like LSP may 
be defined to have $F_g$ very close to 1($\gsim 0.9$). On the other 
side, a Higgsino-like 
LSP would have $F_g \lsim 0.1$. 
For values in between, the LSP could be 
identified as a gaugino-higgsino mixed state.
The MSSM with gaugino masses universal at $M_G$ has a few distinct 
regions in general that satisfy the WMAP constraint.  In this section 
we point out the existence of such regions in the context of the NUSM.
\noindent 
i) The {\it bulk annihilation} 
region in mSUGRA is typically characterized by smaller scalar masses and smaller values of 
$m_{1/2}$ where the LSP is bino-dominated. A bino dominated LSP 
couples favorably with right handed sleptons. Thus, the LSP pair annihilation 
in the bulk annihilation region 
occurs primarily via a t-channel sfermion in mSUGRA. 
There are two important constraints that disfavors the bulk region 
in mSUGRA. These are the constraints a) from the slepton mass lower limit from 
LEP2\cite{limits} and b) from the lower limit of Higgs boson mass $m_h$ of 
$114.4$~GeV\cite{hlim}\footnote{It is quite possible to have a bulk 
region with a non-zero $A_0$\cite{Chattopadhyay:2007di}.}.   
ii) The {\it focus point}\cite{focus} or the 
{\it hyperbolic branch}\cite{hyper}
region of mSUGRA is typically characterized by a small  
$|\mu|$ region that is close to the boundary of the lighter chargino mass 
lower bound.  Because of a small $|\mu|$
here the LSP has a significant amount of the Higgsino component or it can even
be almost a pure Higgsino
\footnote{This holds in the inversion region of the 
Hyperbolic branch\cite{hyper}.}.
Additionally the lighter chargino 
${\tilde \chi}_1^\pm$ becomes lighter and 
coannihilations\cite{Edsjo:1997bg,mizuta} with 
LSP reduce the relic density to an acceptable level.
The HB/FP region is however absent in the NUSM.\\
iii) Coannihilations of LSP may also occur with sleptons, typically 
staus ($ {\tilde \tau}_1$)\cite{coannistau} in mSUGRA. These regions 
are associated with small $m_{1/2}$ and small $m_0$ zones near the 
boundary of the discarded zone where staus become the LSP. 
Stau coannihilation is also an effective way to bring 
the neutralino relic density to an acceptable level in the NUSM. 
Coannihilations of LSP may also occur with stop (${\tilde t}_1$) in a 
general MSSM scenario\cite{coannistop} or even in 
mSUGRA\cite{Chattopadhyay:2007di}.  However, in spite of having 
relatively lighter ${\tilde t}_1$, the NUSM does not have 
such a region unless one reduces the mass of ${\tilde t}_1$ further via 
appropriately considering non-zero values for $A_0$.\\ 
iv) The most important region satisfying WMAP data 
for our study is the Higgs-pole annihilation 
or {\it funnel} 
region\cite{dreesDM93,funnel}. The funnel region that satisfies the WMAP 
data is characterized by the direct-channel pole 
$2m_{{\tilde \chi}_1^0} \simeq m_A,m_H$.  This occurs in 
mSUGRA typically for large $\tan\beta$ extending to larger $m_0$ and 
larger $m_{1/2}$ regions. In the NUSM however, the 
funnel region occurs in all possible 
$\tan\beta (\gsim 5)$. 
 Indeed apart from the LSP-stau coannihilation appearing 
in a small region,  
Higgs-pole annihilation is the primary mechanism in the NUSM to satisfy the 
WMAP constraint throughout the parameter space. 

We now show the results of the computation of the
neutralino relic density using the code micrOMEGAs\cite{micromegas}. 
In Fig.\ref{t10nuspectrum} that corresponds to $\tan\beta=10$ 
the region with red dots in the $(\mhalf-m_0)$ plane satisfy the 
WMAP constraint 
(Eq.\ref{relicdensity}) for the neutralino relic density. 
For a small 
$m_0$ we find the stau coannihilation region marked with 
red dots. The gray region below the stau coannihilation region  
is discarded because of the appearance of charged LSPs. 

The upper gray region is discarded broadly via $m_A^2$ turning negative 
except near the boundary where there can be interplay with other 
constraints as described below.
Thus if we concentrate on the boundary of the discarded region, 
the smallest $m_{1/2}$ zone 
(below 160 GeV or so) is ruled out by the LEP2 lower limit of 
sparticle masses\cite{limits}. This is followed by obtaining 
tachyonic sfermion 
scalars (particularly stop scalars) when $\mhalf$ is increased further up to 
$\simeq$ 600 GeV. The same boundary zone for the larger $m_{1/2}$ region is 
eliminated because of the appearance of the charge and color 
breaking (CCB) minima\cite{casasCCB}.  In the NUSM this happens via the 
CCB conditions that involve $m_{H_D}^2$, the later becoming 
negative makes the CCB constraint stronger. 
In the region 
between the stau coannihilation area and the upper gray shaded discarded area 
one finds a long red region that satisfies the WMAP constraint for 
$\Omega_{{\tilde \chi}_1^0}h^2$.  As mentioned before, the reason for 
satisfying the WMAP data is the direct channel annihilation of two 
LSPs via neutral Higgs bosons.  
Fig.\ref{t10nufunnel} shows a scanned output for 
$\tan\beta=10$ when $m_0$ is varied as in Fig.\ref{t10nuspectrum}. The 
LSP mass is plotted against ${(2\mx1-m_A)}/{2\mx1}$ so as to show the 
extent of the $A$-annihilation or funnel effect. 
Clearly the WMAP satisfied points shown 
in red fall around the zero of the y-axis confirming that the funnel region 
appears in the NUSM for small $\tan\beta$ as well.
We note that the $A$-width can be 
quite large ($\Gamma_A \sim 10$-$50$~GeV) and $2\mx1$
can be appreciably away from the exact 
resonance zone still giving a 
$s$-channel annihilation consistent with the WMAP data. 
The heavy scalar Higgs boson $H$ 
also significantly contributes to the total annihilation cross-section.
\begin{figure}[!ht]
\vspace*{-0.1in}
\mygraph{t10nuspectrum}{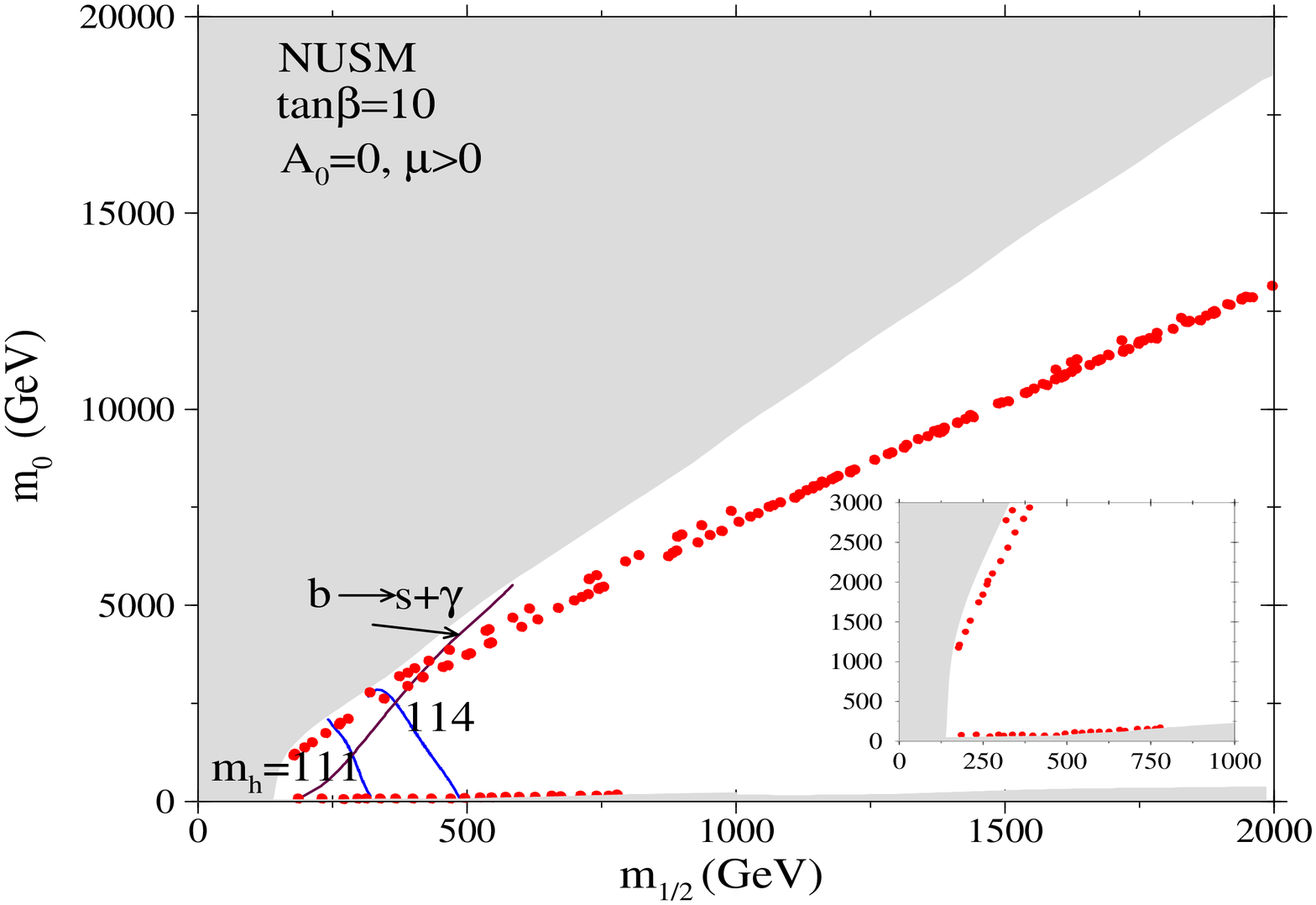}
\hspace*{0.5in}
\mygraph{t10nufunnel}{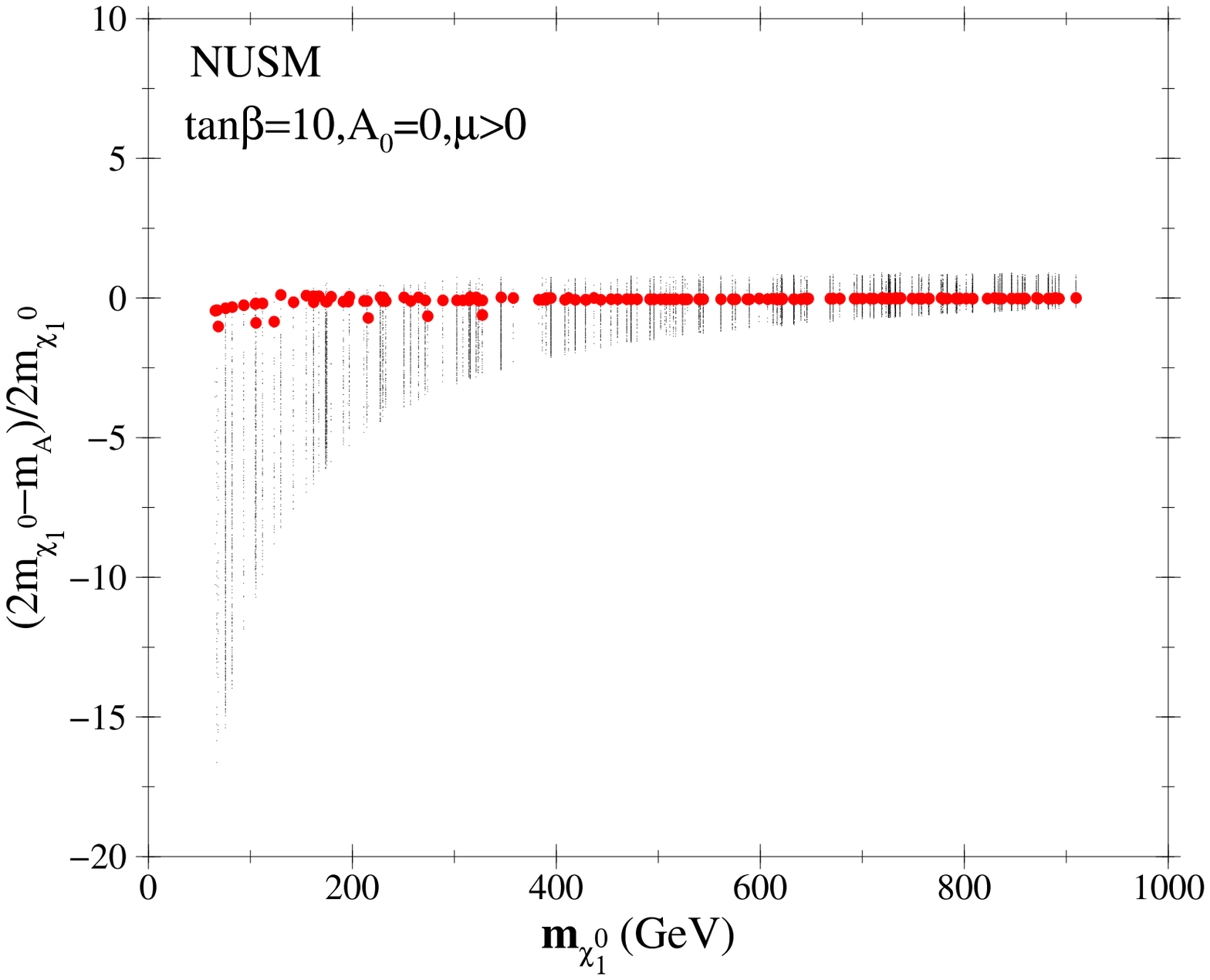}
\caption{
~(a) WMAP allowed region in the $m_{1/2}-m_0$ plane for $\tan\beta=10$ 
and $A_0=0$ with $\mu>0$ for NUSM. 
Lighter Higgs boson mass limits are represented by solid lines. 
Dot-dashed line refers to $b\rightarrow s \gamma$ limit. WMAP
allowed regions are shown by red dots. The 
entire region is allowed via $B_s\rightarrow \mu^+ \mu^-$ bound. The small 
figure within the inset is shown for an improved display of the stau coannihilation zone. ~(b) Scattered points in the plain of $m_{{\tilde \chi}_1^0}$ vs 
$(2m_{{\tilde \chi}_1^0}-m_A)/2m_{{\tilde \chi}_1^0}$ shown 
after a scanning of $m_{1/2}$ and $m_0$ for  
$\tan\beta=10$.   Almost all the WMAP satisfied points (in red) 
occur near the zero of the y-axis, thus suggesting 
the s-channel annihilation of the LSPs via $A$ and $H$ bosons.}
\label{t10dmfigs}
\end{figure}
\begin{figure}[!ht]
\vspace*{-0.1in}
\mygraph{t15nuspectrum}{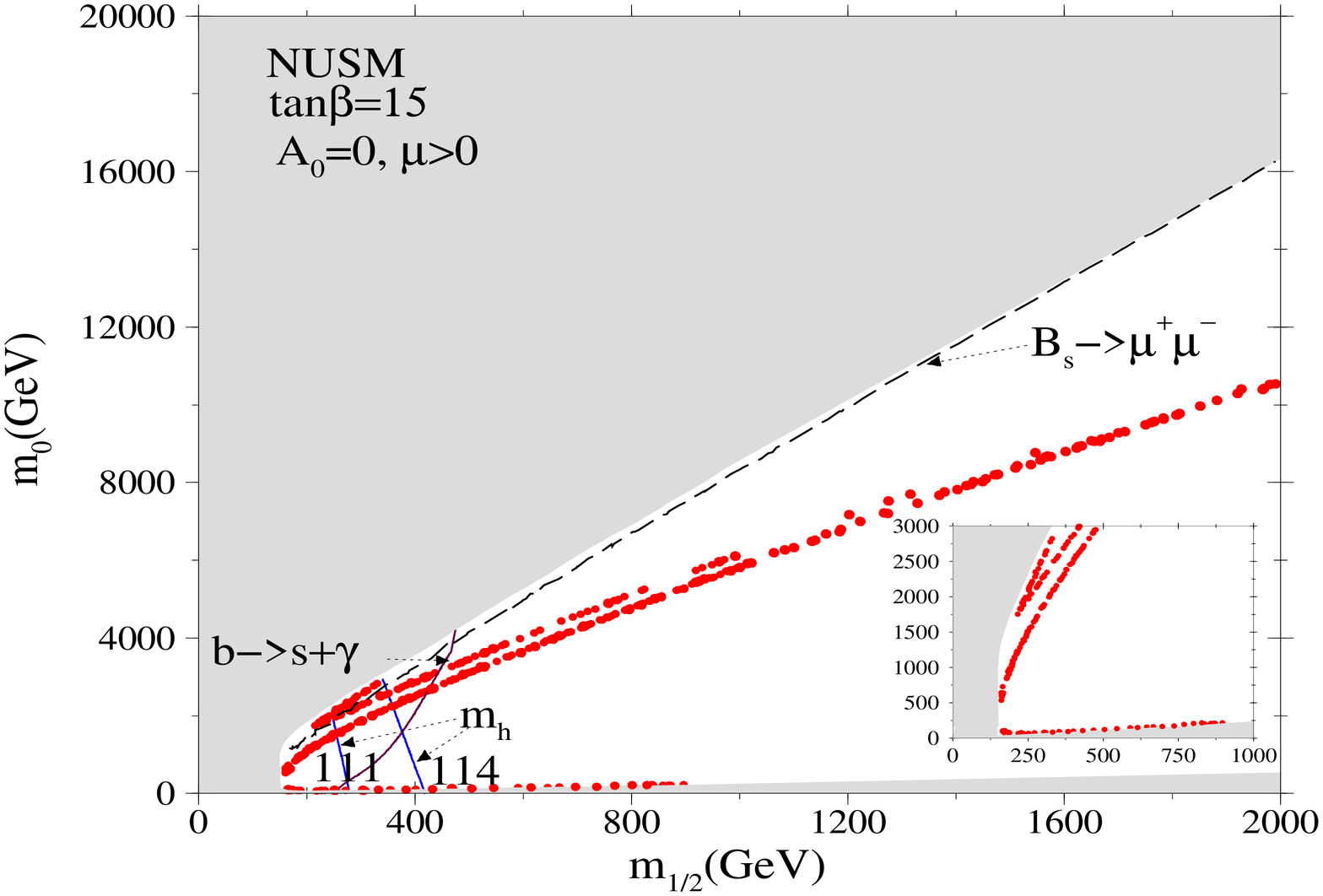}
\hspace*{0.5in}
\mygraph{t15nuhiggsmixing}{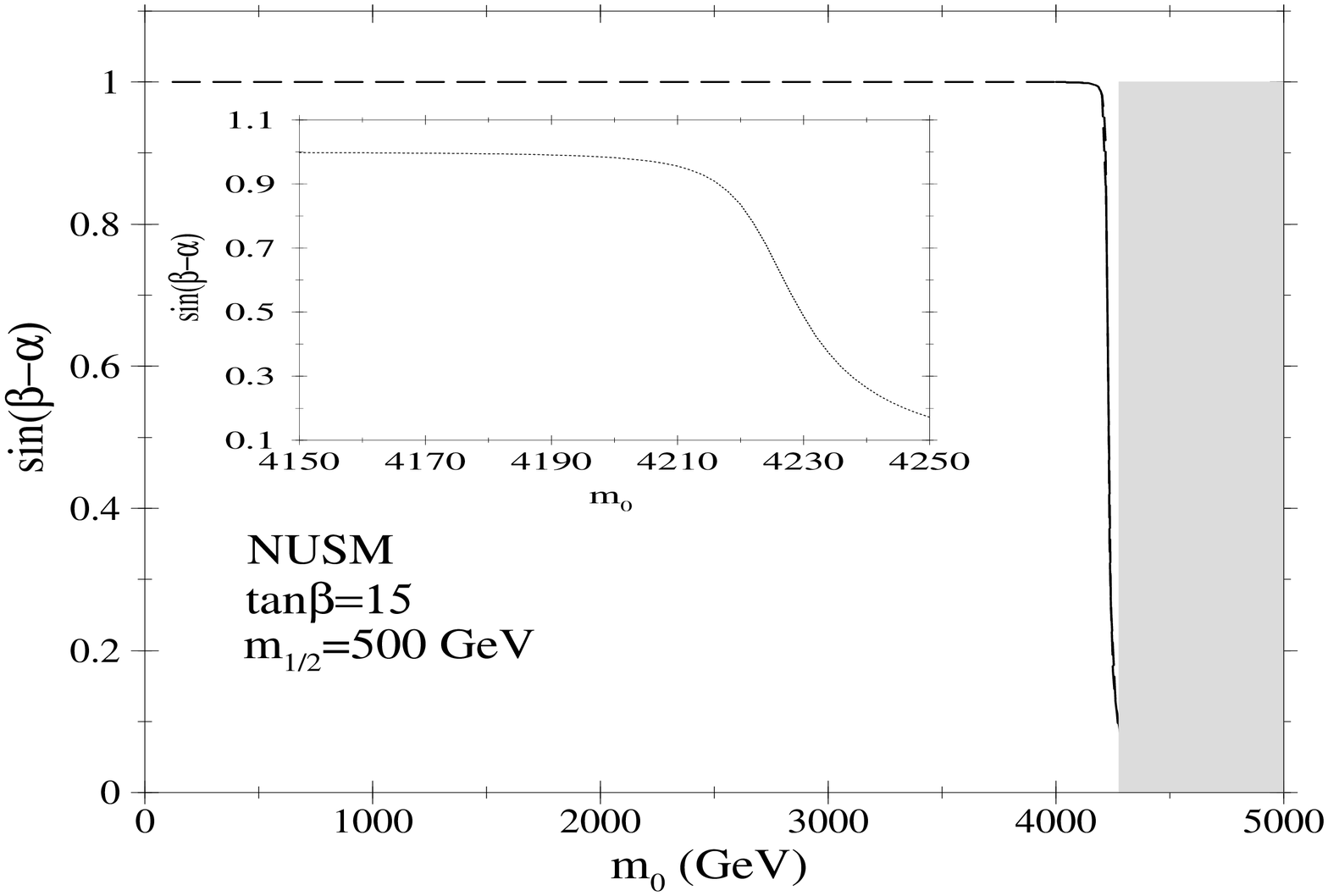}
\caption{
~(a)WMAP allowed region in the $m_{1/2}-m_0$ plane for $\tan\beta=15$ 
and $A_0=0$ with $\mu>0$ for NUSM. 
Lighter Higgs boson mass limits are represented by solid lines. 
Dot-dashed line refers to $b\rightarrow s \gamma$ limit. WMAP
allowed regions are shown by red dots. The 
$B_s\rightarrow \mu^+ \mu^-$ bound is represented by the 
long-dashed line and this discards a small strip of region below 
the discarded top gray region. The small 
figure within the inset is shown for an improved display of the stau 
coannihilation zone. 
~(b) Variation of $\sin(\beta-\alpha)$ vs $m_0$ for $\mhalf=500$~GeV, 
$\tan\beta=15$ and $A_0=0$ for NUSM. The figure 
shows that for large $m_0$, actually near the region where $m_A$ is small, 
$\sin(\beta-\alpha)$ is consistently small. This is of course 
much away from decoupling region of SUSY Higgs boson. The figure in inset 
shows the finer variation of $\sin(\beta-\alpha)$ 
within a small range of $m_0$. The gray shaded region refers to the 
similarly shaded discarded region of Fig.\ref{t15nuspectrum}.  
}
\label{t15dmfigs}
\end{figure}
We now discuss the result of using a few constraints. First, we use the 
LEP2 limit\cite{hlim} of $114.4$ GeV for the SM Higgs boson. For 
the CP-even lighter SUSY Higgs boson mass $m_h$ we use the same constraint 
as long as we are in the decoupling region\cite{decouplingHiggs}  
where the SUSY $h$-boson is  
SM-like. In the NUSM this is true for almost all the parameter space except a 
very small region with small $\mhalf$ and for $\tan\beta \gsim 15$ 
which we will discuss later. Additionally, we 
note that there is an uncertainty of about 3~GeV 
in computing the mass of the light Higgs boson\cite{higgsuncertainty}. This   
theoretical uncertainty primarily originates from momentum-independent 
as well as momentum-dependent two-loop corrections, higher loop corrections 
from the top-stop sector etc. Hence, we have drawn a contour for $m_h=111$~GeV 
in order to consider an effective lower limit.\footnote{The following values 
of top and bottom quarks are considered in the SUSPECT code used in 
our analysis: $m_t=172.7$~GeV and ${m_b}^{{\overline {MS}}}(m_b)=4.25$~GeV.} 
 
We have further drawn the 
$b \rightarrow s \gamma $ contour by considering a 3$\sigma$ 
limit\cite{bsg-recent}, 
\begin{equation} 
2.77 \times 10^{-4} < Br (b \rightarrow s \gamma) < 4.33 \times 10^{-4}. 
\label{bsgammalimits}
\end{equation}
Clearly in Fig.\ref{t10nuspectrum} 
this constraint would keep the $m_{1/2}>425$~GeV region alive and this 
is indeed the region of super-large $m_0 \gsim 4$~TeV 
that satisfies the WMAP data.  
In this context we should however 
note that the $b \rightarrow s+\gamma $ constraint may be of less 
importance 
if some additional theoretical assumptions are taken into consideration. 
The computation of the constraint in models like mSUGRA assumes a 
perfect alignment of the squark and quark mass-matrices. This essentially 
considers an unaltered set of mixing angle factors than 
the Cabibbo-Kobayashi-Maskawa (CKM) factors at the corresponding SM vertices. 
Even a small set of off-diagonal terms in squark mass matrices at the 
unification 
scale may cause a drastic change in the mixing pattern of the 
squark sector at the electroweak scale. This however does not cause any  
effective change in the sparticle mass spectra or in the 
flavor conserving process of neutralino annihilation or  
in generating events in a hadron collider in any significant way. A brief 
review of the model-dependent assumptions for the $b -> s \gamma$ analyses 
may be seen in Refs.(\cite{bsg-okumura}, \cite{djouadidmsugra}) and  
references therein.  

 We now explain an interesting aspect of the NUSM Higgs boson as a consequence 
of the LSM effect. As we have seen before, $m_A$ decreases with increasing 
$m_0$ in the NUSM. We find that with larger 
$m_0$, $m_A$ may become very light   
for the small $\mhalf$ zone so that we may find a spectrum where 
$m_h \sim m_H \sim M_A$ and this indeed 
is a consequence of moving into the intense-coupling\cite{nondecouplingHiggs} 
region of Higgs bosons. This results in a change in the value of the coupling 
$g_{ZZh} \quad (\propto \sin(\beta-\alpha))$, where $\alpha$ is the mixing 
angle between two neutral Higgs bosons $h$ and $H$. 
Typically in the decoupling region $\sin(\beta-\alpha)$ stays very 
close to $1$, but in the intense coupling region this can be considerably 
small like 0.5 or lesser.  This effectively reduces the lower limit of 
$m_h$ to 93 GeV, a value close to $m_Z$. In the NUSM we obtain this effect 
for a small region of parameter space (small $\mhalf$) 
if $\tan\beta \gsim 15$. 
Fig.\ref{t15nuspectrum} shows such a region (this may also satisfy the 
WMAP limits) for 
$200{\rm ~GeV} \lsim \mhalf \lsim 300$~GeV and 
this appears very close to the top gray 
shaded discarded zone. 
Fig.\ref{t15nuhiggsmixing} demonstrates the existence of a small 
$\sin(\beta-\alpha)$ in the NUSM as discussed above.  
However with $A_0=0$, we will see that such a very light 
$m_A$ or $m_h$ region is almost discarded via the present 
limit of the $B_s\rightarrow \mu^+ \mu^-$.
The current experimental limit for the $Br(B_s \to \mu^+ \mu^-)$ 
coming from CDF\cite{CDF} puts
a strong constraint on the MSSM parameter space. 
The experimental bound is given by 
(at ${\rm 95\,\%\,C.L.}$) 
\begin{eqnarray}
{\rm Br} ( B_s \to \mu^+ \mu^-) < 5.8 \times 10^{-8}. 
\label{Bsmumu}
\end{eqnarray}
The estimate of 
$ B_s \to \mu^+ \mu^- $\cite{bsmumu} in the MSSM sensitively depends
on the mass of A-boson ($ \propto m_A^{-4}$) and on the value of 
$\tan \beta$ ($\propto {{\tan}^6\beta}$).  
$ B_s \to \mu^+ \mu^- $ constraint eliminates the thin region along the 
boundary of the REWSB in NUSM where $m_A$ is very light. As mentioned 
before, for $\tan\beta \gsim 15$ a part of the 
above region for small $\mhalf$ zone has 
non-Standard Model like $h$-boson.
 
  Apart from the above constraints we would like to remind that there 
is no non-universality in the first two-generations in NUSM. This saves 
from the stringent FCNC violating limits such as those coming from 
the $K_L$-$K_S$ mass difference or from the 
$\mu \rightarrow e \gamma$ bound. Splitting of the first generation and 
the third generation of scalars may also cause violations of FCNC 
bounds, although to a lesser extent. 
Following 
Ref.\cite{Baer:2004xx} we see that for no violation of FCNC one would 
need (for equal gluino and average squark masses) 
$|m_{\tilde q}(1)-m_{\tilde q}(3)| \lsim m_{\tilde q}^2/M_W$. Here 
$m_{\tilde q}$ refers to the average squark mass. Considering the analysis 
performed in Ref.\cite{Baer:2004xx} for different 
gluino masses we conclude that the results of our analysis stay  
in the safe zone regarding the FCNC bounds in spite of the 
inter-generational splitting between the squarks. 

\noindent
Fig.\ref{t40nuspectrum} shows the results for 
$\tan\beta=40$. Here the funnel 
region is extended up to $m_{1/2}= 1.7$~TeV.  
A large value of $\tan\beta$ increases $h_b$ and $h_\tau$ and 
this would enhance the width $\Gamma_A$\cite{Baer:2003bp}. 
We point out that
unlike Fig.\ref{t10nuspectrum}, here 
the reach of $m_0$ decreases for a given $m_{1/2}$.  The mass of A-boson decreases with 
increase of $m_0$ much more rapidly because of larger $h_b$ and $h_\tau$ 
arising out of larger values of $\tan\beta$. Here, the very light 
$m_A$ region that satisfies the WMAP data and that 
evades the LEP2 Higgs boson limit 
exists near the top gray boundary for 
$250{\rm ~GeV} \lsim \mhalf \lsim 350$~GeV which is again ruled 
out via the $B_s\rightarrow \mu^+ \mu^-$ limit. 
 The upper gray region is typically discarded here via 
the REWSB constraint of $m_A^2$ that turns negative at the tree level. 
Finally, we have not imposed any limit 
from the muon $g-2$ data that may or may not show a discrepancy 
from the Standard Model result. It 
is known that using the recent $e^+$-$e^-$ data leads to a 
$3.4\sigma$ level of discrepancy. On the other hand, 
using hadronic $\tau$-decay 
data in computing the leading order hadronic contribution to muon $g-2$ 
washes away\cite{marciano} any deviation from the SM result.
\begin{figure}[!ht]
\vspace*{-0.1in}
\mygraph{t40nuspectrum}{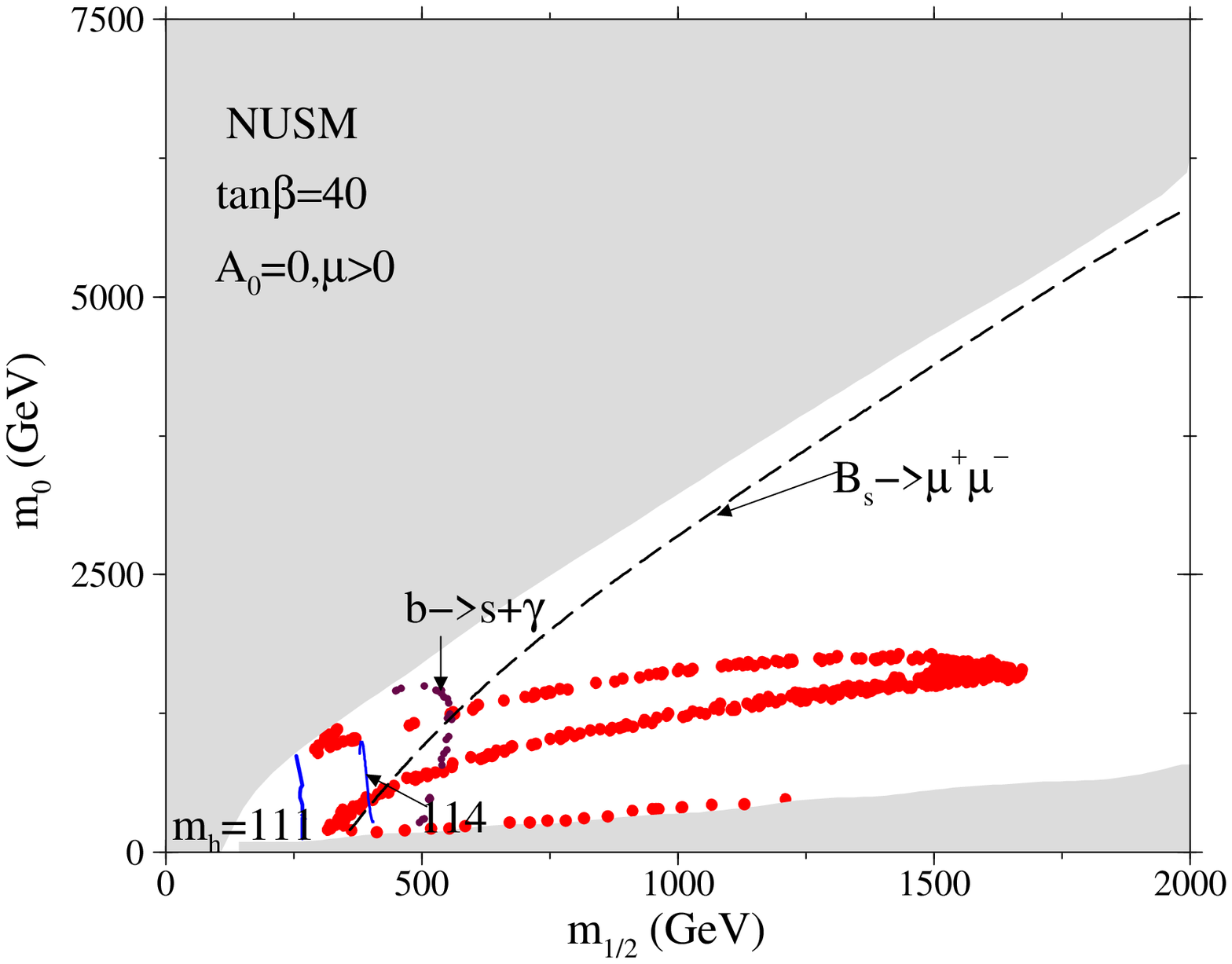}
\hspace*{0.5in}
\mygraph{t40nufunnel}{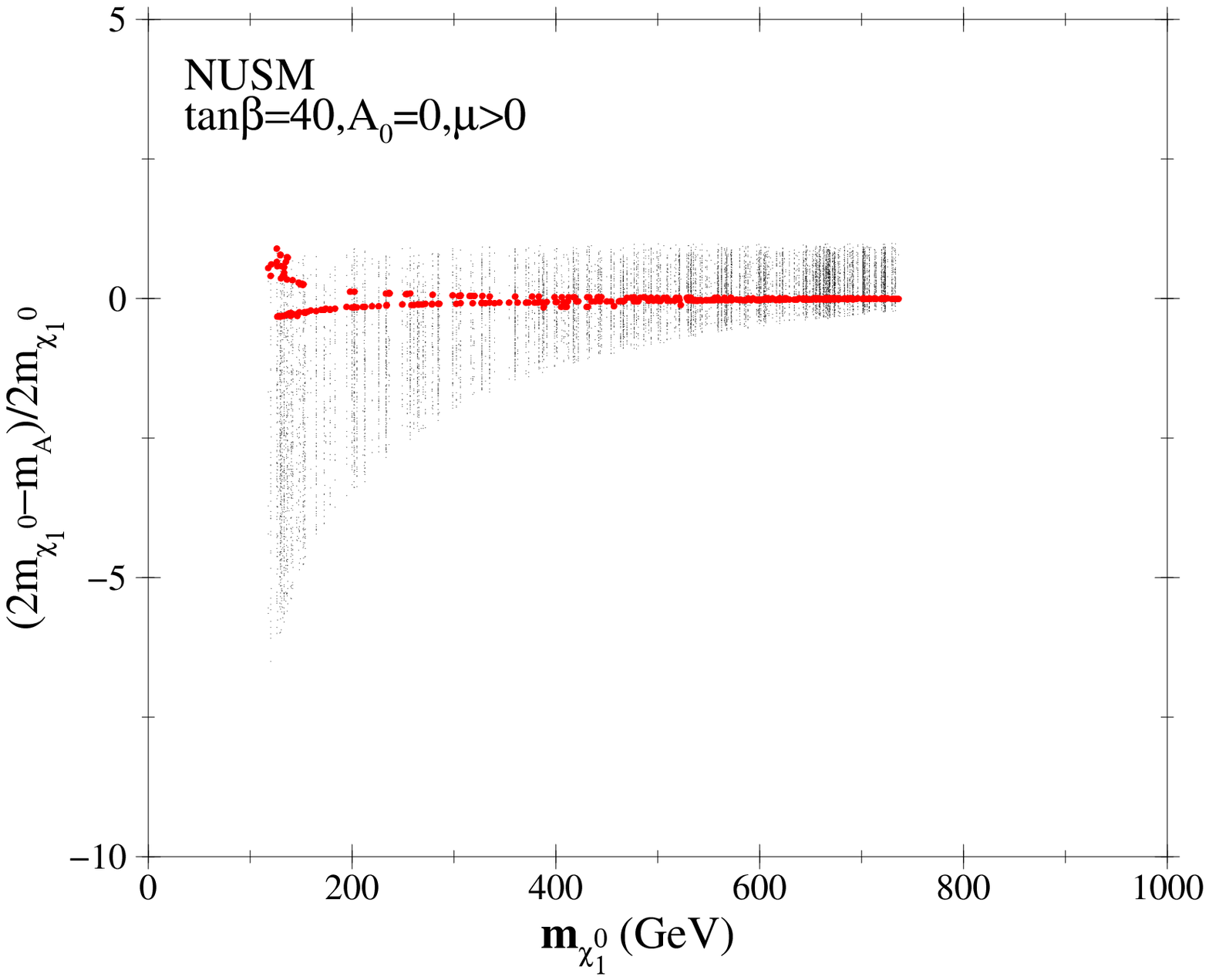}
\caption{
Same as Fig.\ref{t10dmfigs} except $\tan\beta=40$. 
}
\label{t40dmfigs}
\end{figure}

\subsection{Sample parameter points satisfying WMAP data for early 
run of LHC}
We would focus on a few characteristic parameter points to discuss 
the nature of the NUSM spectra that satisfy the WMAP limits. 
Clearly one is able to reach a considerably large $m_0$ satisfying dark matter 
relic density constraints and all other necessary constraints for 
$\mhalf<1$~TeV. As an example with $\mhalf=1$~TeV, this limit is around 
$7$~TeV for $\tan\beta=10$, $6$~TeV for $\tan\beta=15$, and 
$1.6$~TeV for $\tan\beta=40$. Thus, for larger $\tan\beta$ the reach 
of $m_0$ decreases considerably. On the other hand, it is not unusual 
to obtain an $A$-pole annihilation region 
for a large $\tan\beta$ in popular models like 
mSUGRA. Hence, if we are interested in focusing on a region of the MSSM that 
is not available in mSUGRA type of models we would rather 
explore the smaller $\tan\beta$ domain of the NUSM.  
The other important zone of parameter space 
could be the region with smaller $\mhalf$ because this would be easily 
accessible in LHC in its early run. 
We pick up a point on Fig.\ref{t10nuspectrum} ({\it scenario-A}) 
that just satisfies the minimum value of $m_h=111$~GeV besides 
being consistent with the WMAP data for cold dark matter. The input 
parameters are: $\tan\beta=10,A_0=0,\mhalf=270{\rm~GeV},
m_0=2050{\rm~GeV} {\rm~and~} sign(\mu)=1$.  
The scenario-A of Table~\ref{tabnusm} thus has  
light stop and light sbottom quarks, light 
charginos and neutralinos and   
at the same time it would have a light Higgs spectra, all of 
which are promising for an early LHC detection. 
We remind ourselves that typically the NUSM  is associated 
with a heavy first two-generation of scalars 
and heavy sleptons for all the three generations. 
We note that we have relaxed the $b \rightarrow s\gamma$ 
constraint keeping in mind of the argument given 
after Eq.\ref{bsgammalimits}. We could of course respect 
the constraint, only at a price of having a 
little heavier spectra, still that would be very much accessible in LHC for 
the third generation of squarks, Higgs bosons, charginos etc. 
This however would not cause any essential 
change in the general pattern. On the other hand, with not so light 
$m_A$ and with a small $\tan\beta$ 
the scenario-A satisfies the $B_s\rightarrow \mu^+ \mu^-$ limit.
The {\it scenario-B} of Fig.\ref{t15nuspectrum} refers to a special 
point ($\tan\beta=15,A_0=0,\mhalf=255{\rm~GeV}$, $
m_0=2000{\rm~GeV} {\rm~and~} sign(\mu)=1$) for which the Higgs sector 
is not in the decoupling region thus 
reducing the limit of $m_h$ to a value near the Z-boson mass. This parameter 
point obeys the $B_s\rightarrow \mu^+ \mu^-$ limit but 
as with the {\it scenario-A} it has an inadequate 
$Br(b \rightarrow s\gamma)$. With smaller $\mhalf$ and smaller 
$m_0$ {\it scenario-B} has a lighter spectrum in general than  
{\it scenario-A}. Particularly, we find a much smaller Higgs boson 
spectra. 
The {\it scenario-C} (with $\tan\beta=40,A_0=0,\mhalf=540{\rm~GeV}$, $
m_0=1250{\rm~GeV} {\rm~and~} sign(\mu)=1$), that satisfy all the constraints 
including the $b \rightarrow s\gamma$ bound provides a relatively 
heavier spectra.  This is however still within the LHC 
reach for the relevant part of the NUSM  as mentioned above.  We note that 
the large $m_0$ domain of the NUSM  is typically associated with 
lighter Higgs spectrum and this region may in general be probed via 
the production of all MSSM Higgs bosons and subsequently their decays. 
The squarks corresponding to the first two generations and
sleptons of all the generations are 
large when $m_0$ is large.
As a result the production cross-section of these particles will 
be too low at the LHC. Thus, to see any SUSY signal for the NUSM  
one should primarily analyze productions 
and decays of gluino,stop and sbottom in addition 
to charginos and neutralinos.
In a part of the parameter space where the Higgs spectra is light, 
all the Higgs states may be produced 
via $pp \to h,H,A, H^\pm \ +X$
either via loop-induced processes like ($gg \ra h,H,A$) or 
through cascade decays via heavier charginos and neutralinos
$pp \to \chi_2^\pm, \chi_3^0, \chi_4^0
\to  \chi_1^\pm, \chi_2^0, \chi_1^0 + h,H,A, H^\pm $. The latter decays
are only allowed if enough phase space is available.
\begin{center}
\begin{table}[!ht]
\begin{tabular}[ht]{lccc}
\hline
parameter & A & B & C \\
\hline
$\tan\beta$ &10.0  &15.0 &40.0\\
$m_{1/2}$ &270.0  &255.0 &540.0\\
$m_0$ &2050.0  &2000.0 &1250.0\\
$A_0$ & 0  & 0 & 0\\
$sgn(\mu)$ &1  &1 &1\\
$\mu$ & 312.60 & 291.53 &  651.60  \\
$m_{\tg}$ & 709.32 &   674.43 & 1278.26  \\
$m_{\tu_L}$ & 2103.25 &  2047.68 & 1658.64 \\
$m_{\tst_1}$ &  276.89 & 248.27 & 842.00  \\
$m_{\tst_2}$ &   493.53 & 465.21 & 1028.15  \\
$m_{\tb_1}$ & 390.66 & 354.64 &  958.59   \\
$m_{\tb_2}$ &  434.06 & 403.08 & 1019.74 \\
$m_{\te_L}$ & 2050.19 & 1999.70 &1296.33  \\
$m_{{\tilde \tau}_1}$ & 2037.46 & 1972.60 & 1119.32  \\
$m_{{\wt\chi_1}^{\pm}}$ & 196.44 &183.99  & 430.22  \\
$m_{{\wt\chi_2}^{\pm}}$ & 347.34 & 327.11 & 668.46 \\
$\mlspfour$ & 347.72 & 326.97 & 668.18  \\
$\mlspthree$ & 318.11 & 297.66 & 655.13  \\ 
$\mlsptwo$ & 197.69 & 185.15&  430.30 \\ 
$\mlspone$ & 108.05 & 101.72 & 226.91  \\ 
$m_A$ & 259.48 &  148.37 &  403.03  \\
$m_{H^+}$ &  272.027 &  169.44 &  411.73  \\
$m_h$ & 111.26 & 111.25 &116.32 \\
$\Omega_{\tz_1}h^2$& 0.105 & 0.102 & 0.13 \\
$BF(b\to s\gamma)$ & $1.59\times 10^{-4}$ & $4.65\times 10^{-5}$ 
& $2.73\times 10^{-4}$ \\
$BF(B_s\to \mu^+\mu^-)$ & $4.02\times 10^{-9}$ & $2.81\times 10^{-8}$ 
& $5.29\times 10^{-8}$ \\
$\Delta a_\mu    $ & $9.31 \times  10^{-11}$ & $1.59 \times  10^{-10}$ 
& $6.99 \times  10^{-10}$ \\
\hline
\end{tabular}
\caption{Data for point A, B, and C. Masses are in~GeV}
\label{tabnusm}
\end{table}
\end{center}

\section{Direct and Indirect detections of dark matter}
\label{detectionrates}
\subsection{Direct detection rates}
We will now discuss the prospects of direct and indirect 
detections\cite{silkphysrep} 
of neutralino (LSP) as a candidate for dark matter in the NUSM . First, we will discuss 
the direct detection of LSP via measurements of nuclear recoil. 
Neutralinos interact via spin-independent(scalar) and  spin-dependent
interaction\cite{DMreview,directdetection} with nucleons. 
The scalar cross-section may be expressed in terms
of number of protons and neutrons, $Z$ and $(A-Z)$ respectively\cite{DMreview}
as follows.
\begin{equation} 
\sigma_{scalar}= 4 \frac{m^2_r}{\pi} [Zf_p+(A-Z)f_n]^2,
\end{equation}
where $m_r$ is the reduced LSP mass. The quantities $f_p$ and $f_n$ contain 
all the information of short distance physics and nuclear partonic strengths 
and these may be seen in Refs.\cite{directdetdetails}. 
We will now comment on the relative strengths of spin-independent 
and  spin-dependent neutralino-nuclear cross-sections.
While $\sigma_{scalar}$ depends on $Z$ and $A-Z$ quadratically,  
the spin-dependent interaction cross-section on the other hand is 
proportional to $J(J+1)$ where $J$ is the total nuclear 
spin\cite{DMreview}. Typically the spin-independent
neutralino-nucleon scattering cross-sections ( where $\sigma_{\chi p,SI} \simeq 
\sigma_{\chi n,SI}$) are appreciably smaller than the 
corresponding spin-dependent
cross-sections ($\sigma_{\chi p,SD} \simeq 
\sigma_{\chi n,SD}$).  However considering the fact that $\sigma_{SD} \propto 
J(J+1)$ and  $\sigma_{SI} \propto Z^2, (A-Z)^2$, $\sigma_{scalar}$ becomes
considerably larger for moderately heavy 
elements ($A>30$)\cite{Bednyakov:1994qa,KaneKingRev} like Ge, Xe etc. 
But we should keep in mind that 
there exists some cases where $\sigma_{SD}$ may become considerably 
larger than $\sigma_{SI}$ even for 
$A>30$.

\begin{figure}[!ht]
\vspace*{-0.1in}
\mygraph{t10nusmsigma}{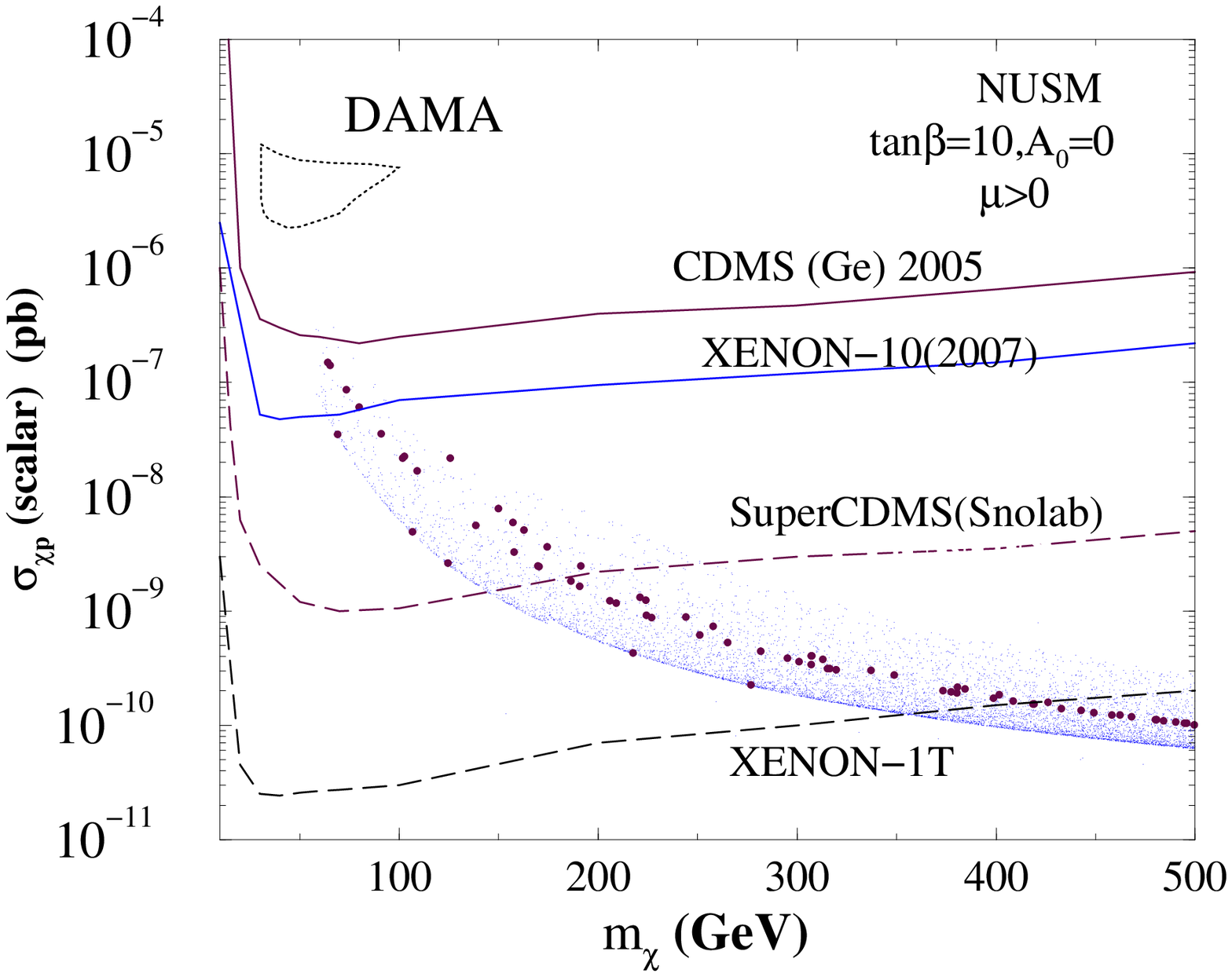}
\hspace*{0.5in}
\mygraph{t40nusmsigma}{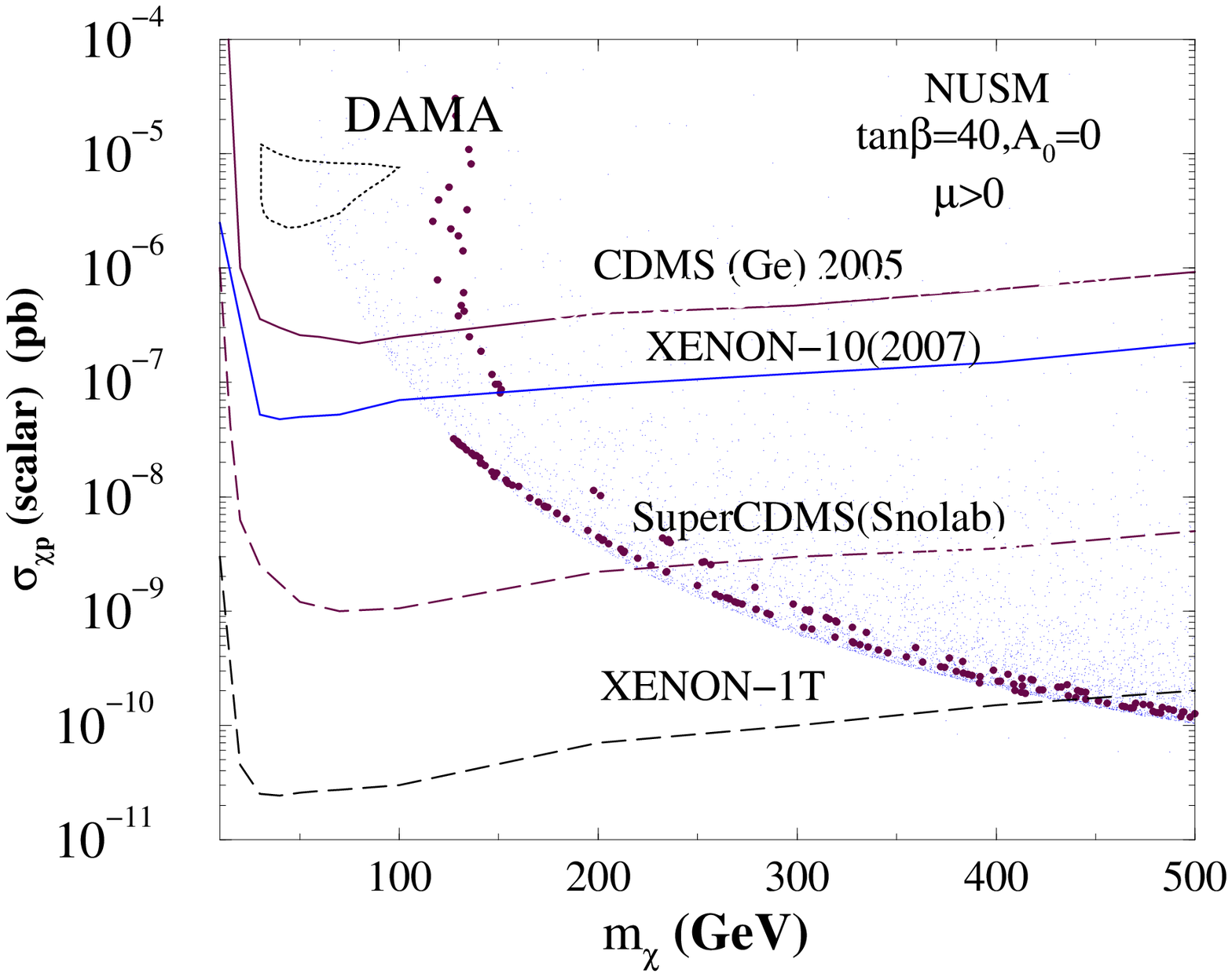}
\caption{Spin-independent scattering LSP-nucleon cross-sections vs LSP mass 
for the NUSM  for $\tan\beta=10$ and 40. The blue dotted region correspond to 
scanned points of the parameter space of Figs.\ref{t10dmfigs} and 
\ref{t40dmfigs}. The maroon filled circles show the WMAP allowed points. 
Various limit plots are shown for different experiments.  
}
\label{scalarxsecs}
\end{figure}

The cross-section $\sigma_{scalar}$ 
mainly involves the computation of $\chi-q$ and
$\chi-\tilde g$ scattering amplitudes. The scalar cross-section at tree
level is composed of $t-$channel Higgs boson exchange and $s-$channel 
squark exchange contributions. 
On the other hand, the spin-dependent cross-section depends on 
$t-$channel $Z$ exchange and $s-$channel squark exchange diagrams.
In this analysis 
we compute the 
spin-independent cross-section $\sigma_{scalar}({\tilde \chi-p})$ 
for two values of $\tan\beta$ ($10$ and $40$) 
for $A_0=0$ and $\mu>0$ using DARKSUSY\cite{darksusy}. 
Fig.~(\ref{scalarxsecs}) shows  
$\sigma_{scalar}({\tilde \chi-p})$ vs $\mx1$ for $\tan\beta=10$ and 
$40$ when $m_{1/2}$ and $m_0$ are varied as in Fig.\ref{t10dmfigs} 
and Fig.\ref{t40dmfigs} respectively. 
The WMAP satisfied points are shown
in small (maroon) circles. We have shown the limits from 
CDMS (Ge) 2005\cite{cdmsGe2005},
XENON-10\cite{xenon102007} and from future experiments 
like SuperCDMS (Snolab)\cite{supercdmssno} 
and XENON1T\cite{xenon1t}. The neutralinos 
with higher mass (up to $400$~GeV) will be partially probed in XENON-1T while
the light LSP region for $\tan\beta=40$ 
is already ruled out by XENON-10 data. 

\subsection{Indirect detection via photon signal}
The fact that the A-resonance annihilation is the primary mechanism 
to satisfy the WMAP limits in the NUSM  suggests that there 
will be enhanced signals for 
indirect detection via $\gamma$-rays, positrons and anti-protons in the NUSM . 
On the other hand, detection via neutrino signal\cite{silkphysrep} 
would not be interesting in 
this case where the LSP is almost a bino annihilating via A-resonance 
in the s-channel\cite{Baer:2003bp}.
Among the above indirect detection possibilities we 
will limit ourselves to estimating only the detection prospect 
of gamma rays that originates from the galactic 
center\cite{gammaraymany1,Bergstrom:1997fj,gammaraymany2,
gammaraymany3}. In general  
for neutralino annihilation at the galactic center one may 
have the following possibilities: i) monochromatic $\gamma$-rays and ii) 
continuum $\gamma$-rays.  Monochromatic $\gamma$-rays
come out from processes like $\chi \chi \rightarrow \gamma 
\gamma$\cite{Ullio1} 
and $\chi \chi \rightarrow Z \gamma$\cite{Ullio:1997ke}. 
These signals although 
small because of the processes being loop suppressed are clean 
with definite energies $E_\gamma=m_\chi$ and 
$E_\gamma=m_\chi-m_Z^2/{4m_\chi}$. Continuum $\gamma$-rays 
on the other hand arise from neutralinos annihilating into a variety of 
Standard Model particles. The
hadronization and production of neutral pions would follow. Typically 
pion decay, in particular $\pi^0 \rightarrow \gamma \gamma$ 
would produce a huge number of photons with varying energies. 

The differential 
continuum $\gamma$-ray flux that arrives from angular direction  
$\psi$ with respect to the galactic center is given 
by\cite{Bergstrom:1997fj,silkphysrep,gammaraymany2}, 
\begin{equation}
\frac{d\Phi_\gamma}{dE_\gamma}(E_\gamma,\psi)=\sum_i \frac{<\sigma_i v>}{8\pi 
m_\chi^2}\frac{dN_\gamma^i}{dE_\gamma} \int_{line~of~sight}ds \rho_\chi^2(r(s,\psi)).
\label{photonfluxeqn}
\end{equation}
Here $\sigma_i$ is a LSP pair annihilation cross section into 
a final channel $i$. We will consider $\gamma$-rays emerging 
from the galactic 
center, hence $\psi=0$. $v$ is the pair's relative velocity and $<\sigma v>$ 
refers to the velocity averaged value 
of $\sigma v$. $\frac{dN_\gamma^i}{dE_\gamma}$
 is the differential $\gamma$-ray yield for the channel $i$. $\rho_\chi (r)
$ is the cold dark matter density at a distance $r$ from the galactic center, where $r^2=s^2+R_0^2-2sR_0\cos\psi$.  Here, $s$ is the line of sight 
coordinate, $R_0$ is the Solar distance to the galactic center. Clearly, 
$\rho_\chi(r)$ that depends on astrophysical 
modelling is important to determine the photon flux in 
Eq.~(\ref{photonfluxeqn}). One may indeed preferably isolate 
the right hand side of Eq.~(\ref{photonfluxeqn}) into a part depending 
on particle physics and a part depending on astrophysics. For the 
later, one defines a dimensionless quantity $J(\psi)$ such that, 

\begin{equation}
J(\psi)=\left(\frac{1}{8.5~{\rm kpc}}\right)
{\left(\frac{1}{0.3~{\rm GeV/cm^3}}\right)}^2\int_{line~of~sight}ds \rho_\chi^2(r(s,\psi)).
\label{jpsieqn}
\end{equation}
The above results in,
\begin{equation}
\frac{d\Phi_\gamma}{dE_\gamma}(E_\gamma,\psi)=
0.94\times 10^{-13} {\rm cm}^{-2}{\rm s}^{-1} {\rm GeV}^{-1} {\rm sr}^{-1}
\sum_i \frac{dN_\gamma^i}{dE_\gamma}\left(\frac{<\sigma_i v>}{10^{-29} 
{\rm cm}^3 {\rm s}^{-1}}\right) {\left(\frac{100~ {\rm GeV}}{m_\chi}\right)}^2 
J(\psi).
\label{newphotonfluxeqn}
\end{equation}
For a detector that has an angular acceptance $\Delta \Omega$ and lowest 
energy threshold of $E_{th}$ the total gamma ray flux from the galactic 
center is given by,
\begin{equation}
\Phi_\gamma(E_{th})=0.94\times 10^{-13} {\rm cm}^{-2}{\rm s}^{-1} \sum_i
\int_{E_{th}}^{m_\chi} dE_\gamma \frac{dN_\gamma^i}{dE_\gamma}\left(\frac{<\sigma_i v>}{10^{-29} 
{\rm cm}^3 {\rm s}^{-1}}\right) {\left(\frac{100~ {\rm GeV}}{m_\chi}\right)}^2 
{\bar J}(\Delta \Omega) \Delta \Omega .
\label{totalphotonflux}
\end{equation}
Here ${\bar J}(\Delta \Omega)=\frac{1}{\Delta \Omega}\int_{\Delta \Omega}
J(\psi)d\Omega $. 
The upper limit of the integral in Eq.(\ref{totalphotonflux}) 
is fixed by the fact that the neutralinos move 
with galactic velocity, therefore the annihilations may be considered 
to have occured at rest. 
We will now comment on the galactic halo density profiles used in this 
analysis. Various N-body simulations suggest that one may obtain a 
general profile behavior arbitrary to the extent of a few parameters and 
this is given by\cite{Hernquist:1990be}, 
\begin{equation}
\rho(r)=\rho_0
{{\left[1+{(R_0/a)}^\alpha\right]^{\frac{(\beta-\gamma)}{\alpha}}} \over
{{(r/R_0)}^\gamma{\left[1+{(r/a)}^\alpha\right]}^
{\frac{(\beta-\gamma)}{\alpha}}}}.
\label{genprofile}
\end{equation}
Here $\rho_0$ is a normalisation factor which is taken as the local 
({\em i.e.} solar region) halo density ($\simeq 0.3~{\rm GeV}/{\rm cm}^3$). We will 
analyze with  
three popularly used 
profiles, namely the isothermal cored\cite{isothermalcore}, 
Navarro, Frenk and White (NFW) profile\cite{NFW} and 
Moore profile\cite{Moore} as given in the Table~(\ref{profiletable}).  
The table also mentions the corresponding value of $\bar J$ for 
$\Delta \Omega=10^{-3}$ and $10^{-5}$ sr. Computation of the photon flux 
for a different 
halo profile may easily be performed by an appropriate scaling with 
the corresponding $\bar J$. Clearly, more 
cuspy profiles would produce higher photon-flux. One can further include 
the effects of baryons on the dark matter halo profiles. Baryons may undergo 
radiative processes leading to a fall towards the central region of a 
galaxy in formation.  This changes the density profiles of matter towards 
the center which in turn leads to an increased concentration of dark matter. 
Adiabatic compression\cite{Blumenthal:1985qy} has been used
to study the baryonic effects.  
Inclusion of the adiabatic compression effects cause the profiles to become 
significantly cuspier, often increasing $\bar J$ by a factor of 
100 or so\cite{gammaraymany2}.  
We have not included these halo profile models in our 
computation, but the photon flux would increase by a similar factor as 
mentioned above.
\begin{center}
\begin{table}
\begin{tabular}[t]{|l|l|l|l|l|l|l|l|}
\hline
Halo Model & $a$~(kpc)  & $R_0$~(kpc) & $\alpha$ & $\beta$ & $\gamma$ & $\bar J(10^{-3})$ 
& $\bar J(10^{-5})$ \\
\hline 
\hline
Isothermal cored & 3.5 & 8.5 & 2 & 2 & 0 & 30.35 & 30.4 \\ 
\hline
NFW & 20.0 & 8.0 & 1 & 3 & 1 & $1.21 \times 10^3$ & $1.26 \times 10^4$ \\ 
\hline
Moore & 28.0 & 8.0 & 1.5 & 3 & 1.5 & $1.05 \times 10^5$ & $9.75\times 10^6$ \\
\hline
\end{tabular}
\caption{A few dark matter halo density profiles and associated parameters.}
\label{profiletable}
\end{table}
\end{center}
\begin{figure}[!ht]
\vspace*{-0.1in}
\mygraph{t10mh400photon}{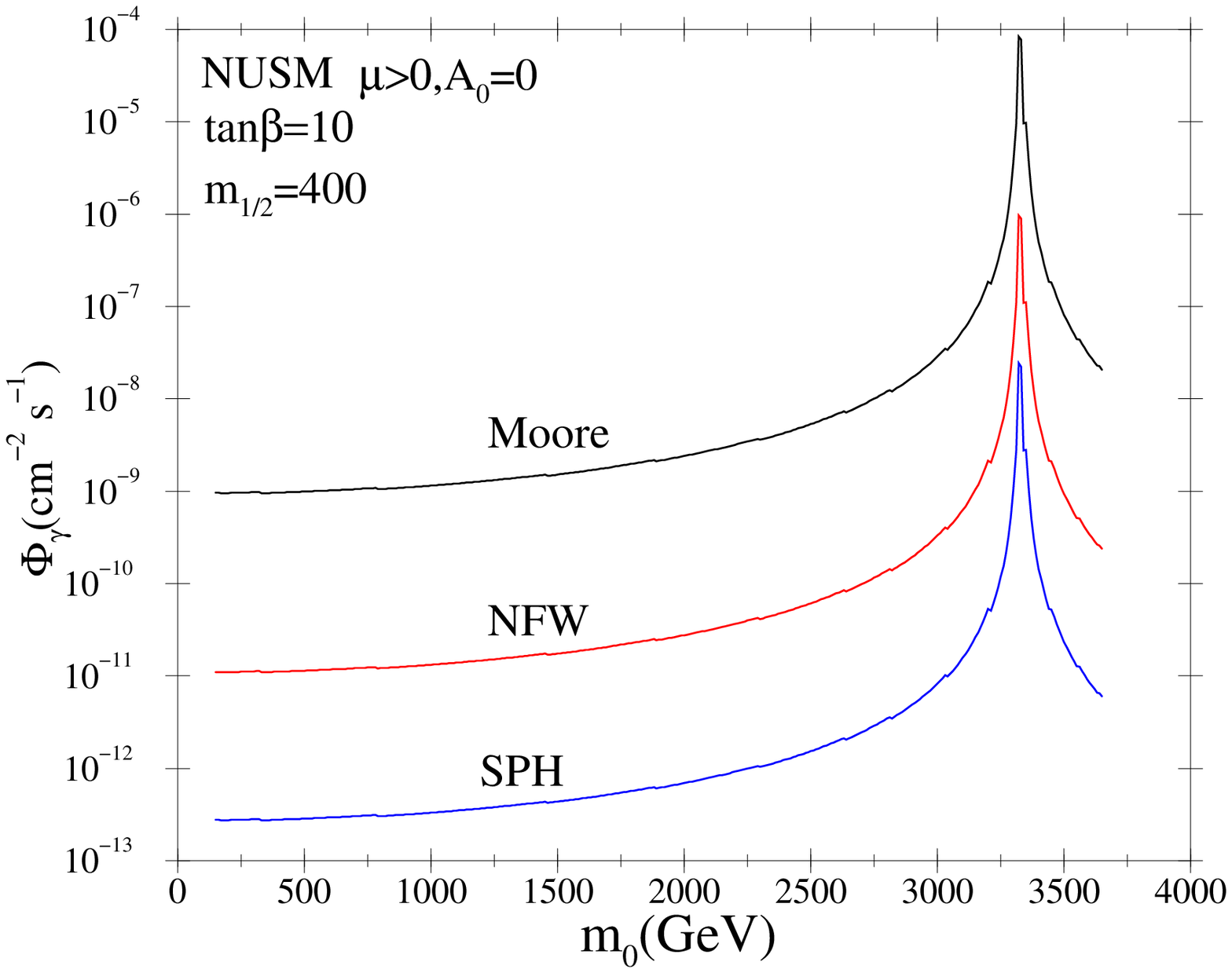}
\hspace*{0.5in}
\mygraph{t10mh800photon}{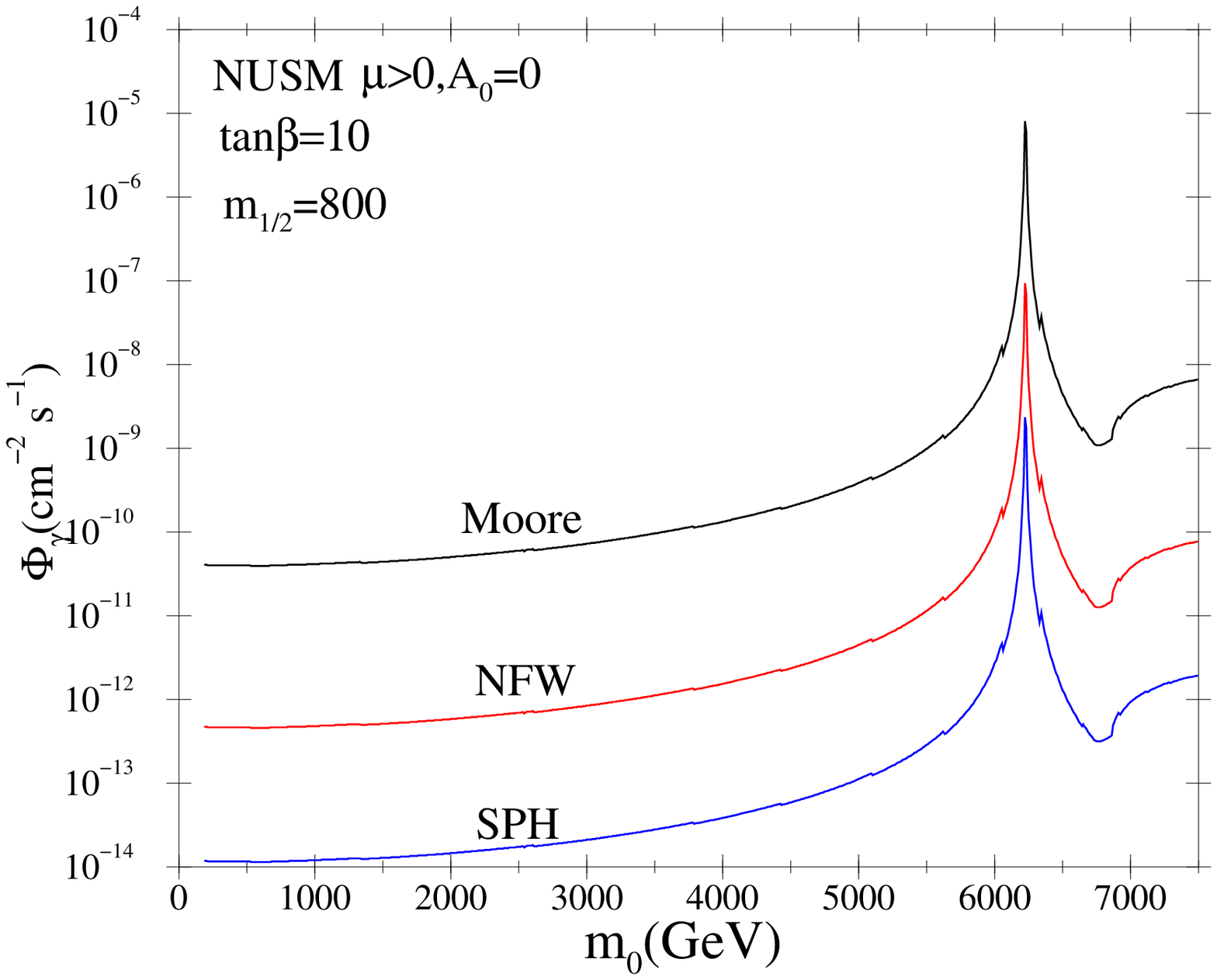}

\vspace*{0.5cm}
\mygraph{t40mh400photon}{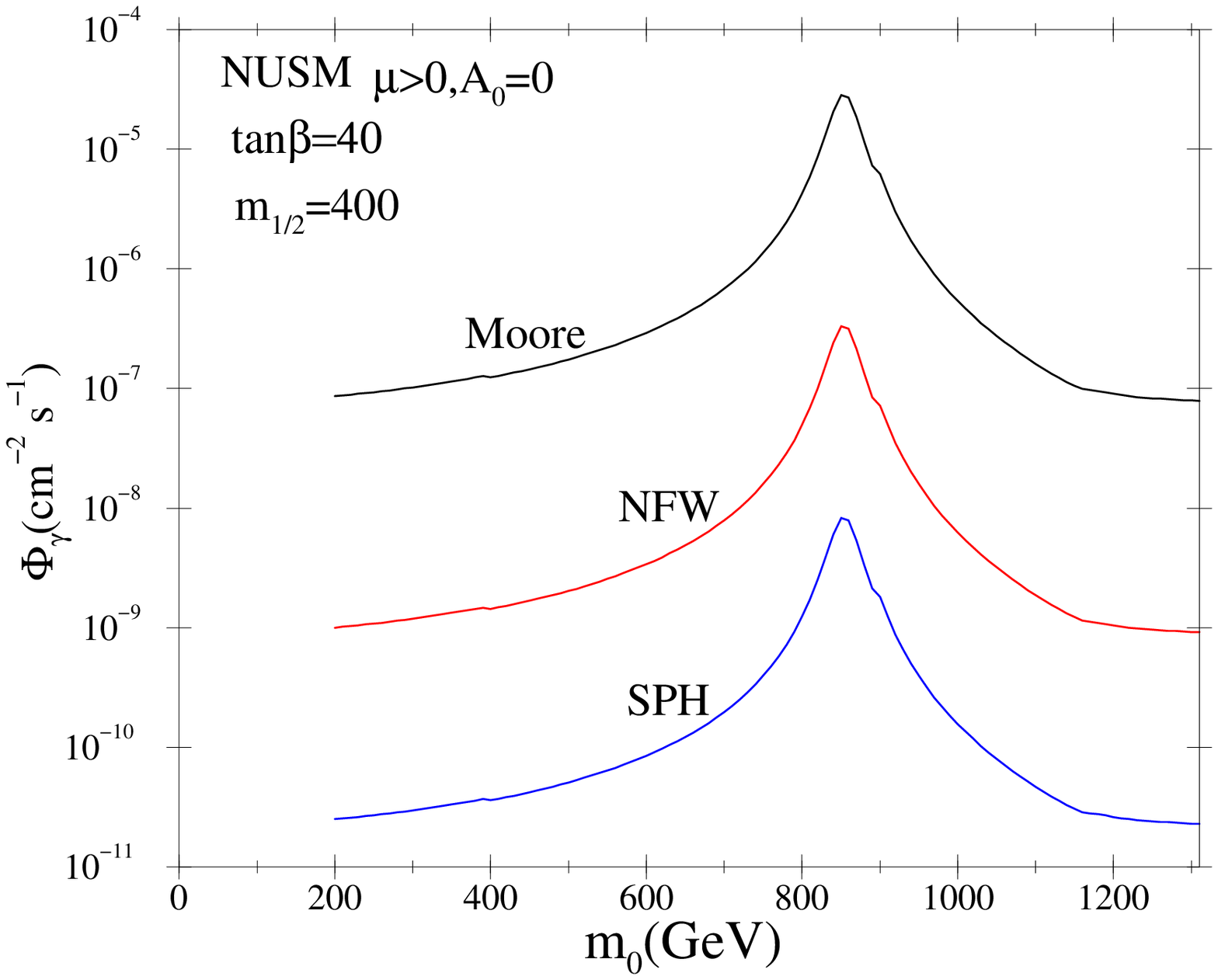}
\hspace*{0.5in}
\mygraph{t40mh800photon}{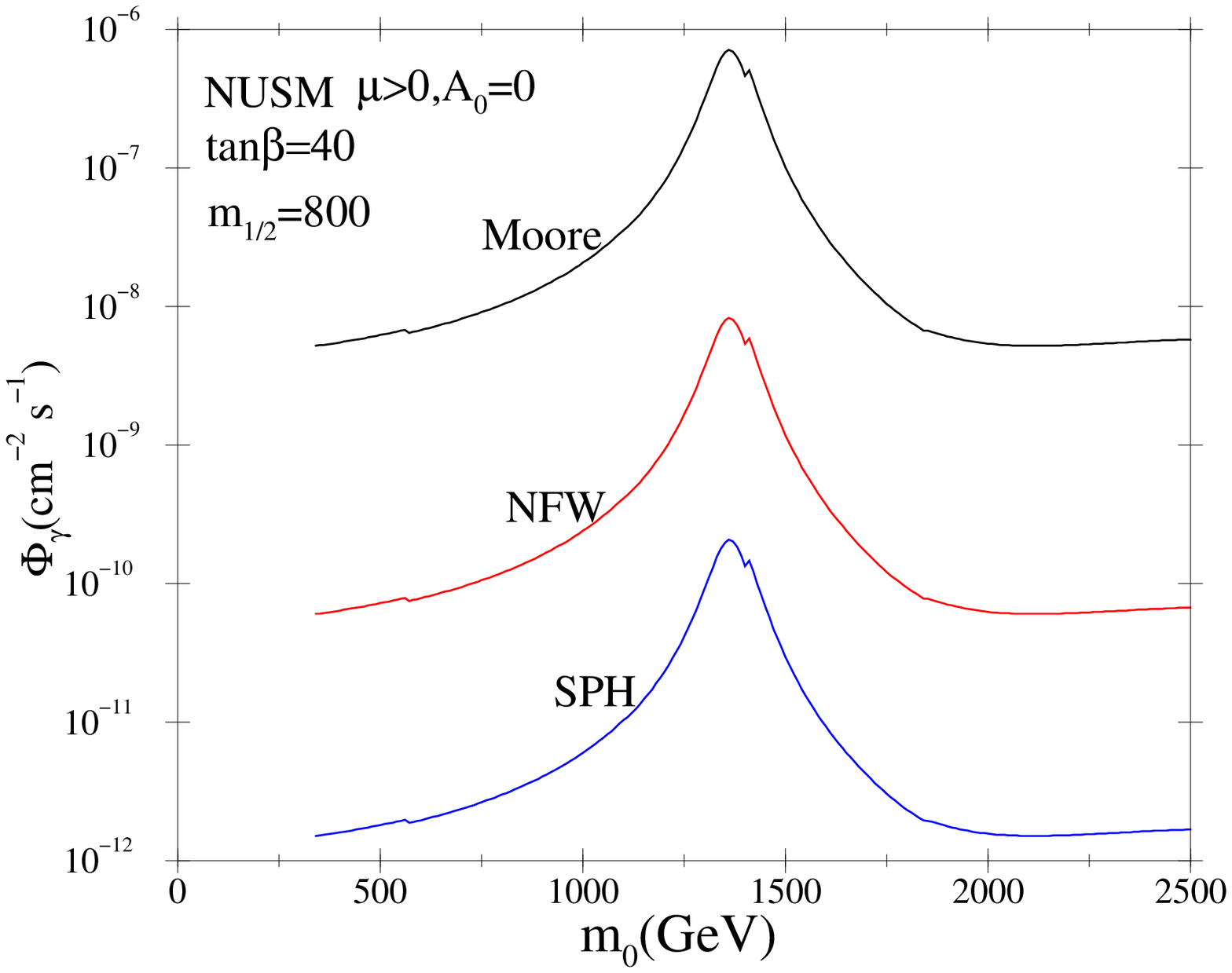}
\caption{Continuous $\gamma$-ray flux in ${\rm cm}^{-2} {\rm s}^{-1}$ 
above a threshold energy of 
$1$ GeV for a cone of $1 \times 10^{-3}$~sr centered around the 
galactic center {\em vs} $m_0$. Lines are shown for three 
different halo distributions i) spherically symmetric 
isothermal cored profile 
(SPH)\cite{isothermalcore}, ii) Navarro, Frenk and White (NFW) 
profile\cite{NFW} and iii) Moore profile\cite{Moore}. 
}
\label{photonfluxvsm0}
\end{figure}
Fig.(\ref{photonfluxvsm0}) shows the result of 
continuum photon-flux {\em vs} $m_0$ 
for $\tan\beta=10$ and $40$ corresponding to two different values of 
$m_{1/2}$~($=400$ and $800$~GeV) 
and three different halo profiles as mentioned above. The photon flux 
(in ${\rm cm}^{-2} {\rm s}^{-1}$) in the NUSM  is computed using DARKSUSY\cite{darksusy} 
with $E_\gamma>1$~GeV, for a solid angle aperture of 
$\Delta \Omega=10^{-3}$~sr.    
As $m_0$ increases 
the mass of psuedoscalar Higgs boson decreases and the LSP 
pair annihilation through s-channel resonance 
causes a peak corresponding to a halo profile. Broadly, such a 
peak covers the region of $m_0$ for a given $m_{1/2}$ 
where WMAP data for the neutralino relic density is satisfied. With 
an increase in $\tan\beta$ the width $\Gamma_A$ of $m_A$ increases 
and the peak associated with a given halo profile broadens. We see that in spite 
of having a broad range of halo profile characteristics, the photon flux 
in the region of resonance annihilation in the NUSM  where the WMAP data 
is satisfied may be probed in the upcoming GLAST\cite{glast1,glast2} 
experiment at 
least for the cuspier profiles. 
GLAST would be able to probe photon-flux as low as $10^{-10}$ ${\rm photons}/ {\rm cm}^2/ 
{\rm s}$\cite{glast2}.  
We note that this conclusion remains valid in spite of the 
fact that GLAST will use an aperture of $\Delta \Omega=10^{-5}$~sr so that 
an appropriate scaling of the photon-flux in Fig.(\ref{photonfluxvsm0}) 
needs to be done from Table~(\ref{profiletable}) and 
Eq.(\ref{totalphotonflux}). We further note that as mentioned 
before, the use of the adiabatic 
compression mechanism would modify a given halo profile 
significantly and this may increase the photon flux by a few orders of  
magnitude.
\begin{figure}[!ht]
\vspace*{-0.1in}
\mygraph{t10photonvsmchi}{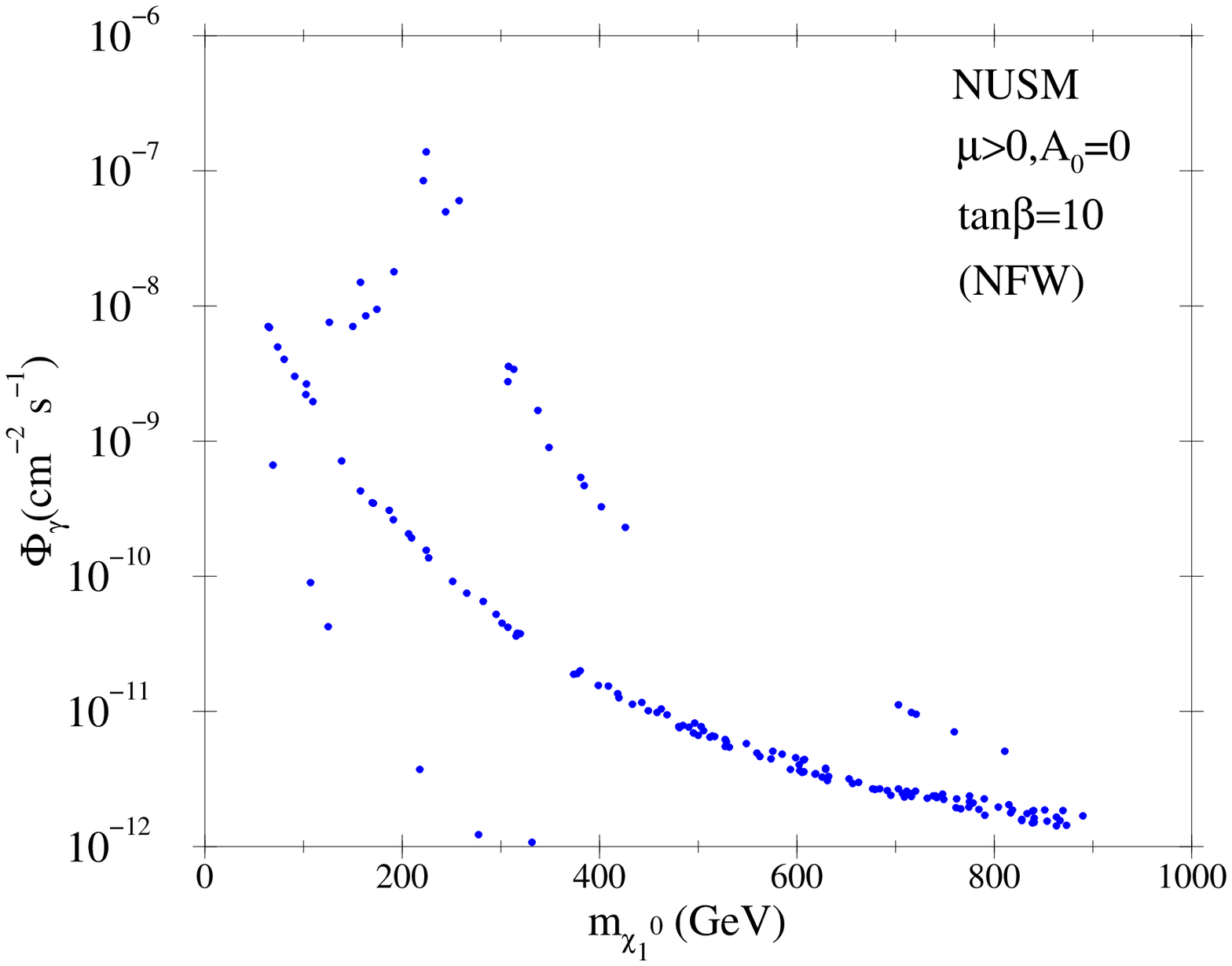}
\hspace*{0.5in}
\mygraph{t40photonvsmchi}{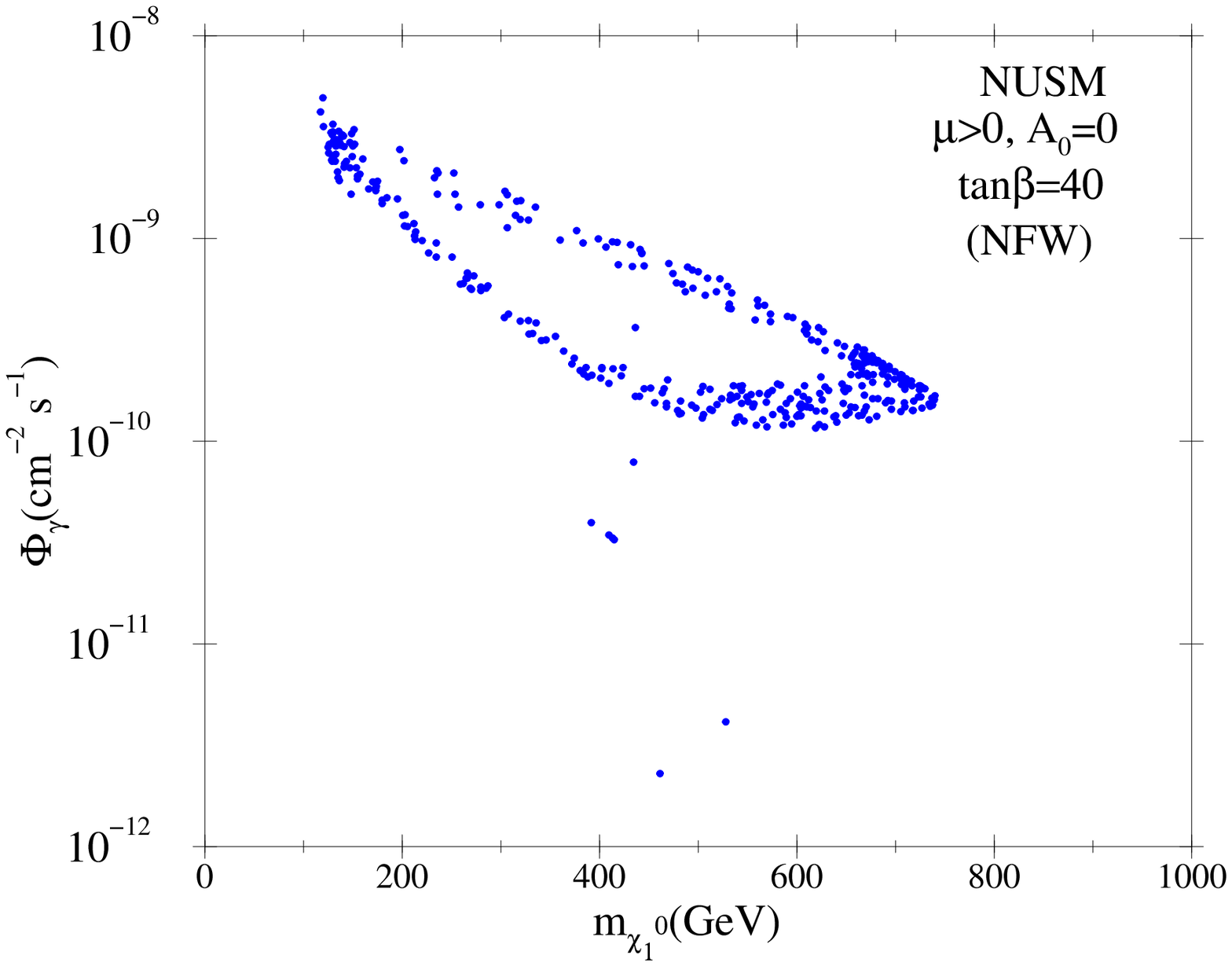}
\caption{Scatter plot of photon flux 
in ${\rm cm}^{-2} {\rm s}^{-1}$ (with $E_\gamma>1$~GeV) 
vs LSP mass for $\tan\beta=10$ and $40$ for a NFW halo profile in the NUSM .  
Here $m_{1/2}$ and $m_0$ are scanned in the 
ranges shown in Figs.(\ref{t10dmfigs},\ref{t40dmfigs}). Only WMAP relic 
density satisfied points are shown. 
}
\label{photonfluxvsmchi}
\end{figure}
Fig.(\ref{photonfluxvsmchi}) shows the plots of photon-flux 
(in ${\rm cm}^{-2} {\rm s}^{-1}$ with $E_\gamma>1$~GeV) vs the mass of 
the LSP for $\tan\beta=10$ and $40$ for 
NFW halo profile in the NUSM . Here, $m_{1/2}$ and 
$m_0$ are varied such that $m_{1/2}<2$~TeV and $m_0<20$~TeV for $\tan\beta=10$ 
and $m_{1/2}<2$~TeV and $m_0<10$~TeV for $\tan\beta=40$. Only 
WMAP allowed parameter points are shown. GLAST would be able to probe 
up to $400$ to $450$ GeV. 

\section{Conclusion}
\label{concl}
In this analysis we have worked with a non-universal scalar mass 
scenario in a supergravity framework. We started with purely 
phenomenological motivations namely, 
i) to manage the FCNC and CP-violation type of 
constraints by decoupling, ii) to obtain WMAP satisfied values for 
neutralino relic density for a 
broad region of parameter space without depending on any delicate 
mixing of gauginos and Higgsinos, iii) to have radiative electroweak 
symmetry breaking and 
iv) to keep naturalness within control. Keeping the above in mind and 
considering a unified 
gaugino mass scenario, we used a common scalar mass parameter $m_0$ 
at the gauge coupling unification scale for 
the first two-generation of scalars as well as the third generation of 
sleptons. The item (i) mentioned above would require $m_0$ to be large, and 
the item (iv) would prefer light third generation of squarks and light 
Higgs scalars. For simplicity we used vanishing third 
generation of squarks and Higgs scalar masses at the unification scale. 
In such a scenario with 
a possibly multi-TeV $m_0$, 
we first found a large mass effect or in particular 
large slepton mass effect in the RGE of  
$m_{H_D}^2$ ({\em ie.} for the down type of Higgs scalar) that turns the later 
negative at the electroweak scale 
almost irrespective of a value of $\tan\beta$. Extending the semi-analytic 
solution of $m_A^2$ by  
considering the $h_b$ and the $h_\tau$ terms (this is required here 
even for a small $\tan\beta$) relevant for the large 
slepton mass effect we found a hyperbolic branch/focus 
point like effect in $m_A^2~(\simeq m_{H_D}^2 - m_{H_U}^2)$ 
for small $\tan\beta$. This causes $m_A$ to be almost 
independent of $m_0$ for a large domain of the 
later. But, with a very large $m_0$ this causes $m_A$ to become 
very light or this may even turn $m_A^2$ negative 
giving rise to no radiative  
electroweak symmetry breaking. We further found that because of such 
large slepton mass effect in the RGE, the 
Higgs sector may reach an intense coupling region with all the Higgs 
bosons becoming very light and this may evade the LEP2 limit of 
$m_h$ for a limited region of parameter space. However, 
constraint from $Br(B_s\rightarrow \mu^+ \mu^-)$ becomes stringent 
in this region because of lighter $m_A$.  In general, 
we find relatively lighter $m_A$ or $m_H$ and this fact 
leads to a large 
$A$-pole annihilation region or funnel region of dark matter 
even for a small $\tan\beta$. This is in contrast to minimal supergravity 
type of scenarios where funnel region may occur only for large 
values of $\tan\beta$. 
The nature of the LSP is bino-dominated, thus 
there is no need of any delicate mixing of binos and Higgsinos in order 
to satisfy the neutralino relic density constraint. We have also computed 
the direct detection rates of LSP-nucleon scattering. The upcoming 
detectors like XENON-1T would be able to probe almost the entire region of  
parameter space. We have further estimated the 
indirect detection prospect 
via computing continuous photon fluxes. The ongoing GLAST experiment 
will be succesfully able to probe the parameter space even for a less cuspy 
halo profile. We have also briefly discussed the 
detection prospect of sparticles in the LHC. The  Higgs bosons and  the 
third generation of squarks are
light in this scenario. In addition to charginos and neutralinos 
the above may be  easily probed in the early runs of LHC.
\section{Appendix}
\label{append}
The coefficients appearing in Eqs.~\ref{mueqn} and \ref{mAeqn} 
are given by,

\begin{equation}
C_2 =\frac{\tan^2\beta}{(\tan^2\beta-1)}k, \quad 
C_3 =-\frac{1}{(\tan^2\beta-1)}(g-e\tan^2\beta), ~{\rm ~and}~
C_4 =-\frac{\tan^2\beta}{(\tan^2\beta-1)}f 
\end{equation}

\begin{equation}
D_2 =\frac{\tan^2\beta+1}{\tan^2\beta-1} k,  \quad 
D_3 =-\frac{\tan^2\beta+1}{\tan^2\beta-1} (g-e),   ~{\rm ~and}~
D_4=-\frac{\tan^2\beta+1}{\tan^2\beta-1}f 
\end{equation}

Here the functions $k,g,e$ and 
$f$ may be seen in Ref.~\cite{Ibanez:1984-85}. 
Electroweak scale result of $Y_i=h_i^2/{(4\pi)}^2$ with $i\equiv t,b,~{\rm 
and}~\tau$ are shown below. 
\beq
Y_1(t) = \frac{E_1(t) Y_1(0)} {1 + 6 Y_1(0) F(t)}, \quad
Y_2(t) = \frac{E_2(t) Y_2(0)} {(1 + 6 Y_1(0) F(t))^\frac{1}{6}}, \quad
{\rm and}~Y_3(t) = Y_3(0)E_3(t)
\label{yukawa}
\eeq
The quantities $E_i(t)$ are defined as follows. 
\begin{eqnarray}
E_1(t)&=&(1+\beta_3(t))^{\frac{16}{3b_3}}(1+\beta_2(t))^{\frac{3}{b_2}}
(1+\beta_1(t))^{\frac{13}{9b_1}}\nonumber \\
E_2(t)&=&(1+\beta_1(t))^{\frac{-2}{3b_1}} E_1(t)\nonumber \\
E_3(t)&=&(1+\beta_1(t))^{\frac{3}{b_1}}(1+\beta_2(t))^{\frac{3}{b_2}}
\end{eqnarray}
Here, 
$F(t)=\int_0^t E_1(t^\prime)dt^\prime$, $\beta_i=\alpha_i(0)b_i/4\pi$ and 
$(b_1,b_2,b_3)=(33/5,1,-3)$.

\noindent
{\bf Acknowledgments}\\ 
DD would like to thank the Council of Scientific
and Industrial Research, Govt. of India for the support received as a
Junior Research Fellow. UC is thankful to the organizers of the workshop 
``TeV Scale and Dark Matter (2008)'' at NORDITA where the last part of 
the work was completed. UC acknowledges useful discussions with  
M. Guchait, A. Kundu, B. Mukhopadhyaya, D. Choudhury, 
D.P. Roy, Jan Kalinowski and S. Roy.

\end{document}